\documentclass[fleqn,usenatbib]{mnras}
\usepackage[T1]{fontenc}
\usepackage{ae,aecompl}
\usepackage[FIGTOPCAP,nooneline]{subfigure}
\usepackage{comment} 
\usepackage{newtxtext,newtxmath} 
\newsavebox{\circlebox}
\savebox{\circlebox}{\fontencoding{OMS}\selectfont\large\char13}
\newlength{\circleboxwdht}
\newcommand{\ctext}[1]{%
  \setlength{\circleboxwdht}{\wd\circlebox}%
  \addtolength{\circleboxwdht}{\dp\circlebox}%
  \raisebox{0.4\dp\circlebox}{%
    \parbox[][\circleboxwdht][c]{\wd\circlebox}{\centering\small #1}}%
  \llap{\usebox{\circlebox}}%
}

\usepackage{graphicx}	% Including figure files
\usepackage{amsmath}	% Advanced maths commands
\usepackage{amssymb}	% Extra maths symbols

\title[Segmented Telescope PSF by Symmetric Formulation]{Point Spread Function of Hexagonally Segmented Telescopes by New Symmetrical Formulation}
 
\author[S. Itoh et al.]{
S. Itoh,$^{1}$\thanks{E-mail: sitoh@iral.ess.sci.osaka-u.ac.jp}
T. Matsuo,$^{1}$
H. Shibai,$^{1}$
T. Sumi$^{1}$
\\
$^{1}$Osaka university, Faculty of earth and space science, 1-1, Machikaneyama-cho, Toyonaka, Osaka, Japan, 560-0043
}

\date{Accepted XXX. Received YYY; in original form ZZZ}

\pubyear{2018}

\begin{document}
\label{firstpage}
\pagerange{\pageref{firstpage}--\pageref{lastpage}}
\maketitle
  
\begin{abstract}
A point spread function of hexagonally segmented telescopes is derived by a new symmetrical formulation. By introducing three variables on a pupil plane, the Fourier transform of pupil functions is derived by a three-dimensional Fourier transform. The permutations of three variables correspond to those of a regular triangle's vertices on the pupil plane. 
The resultant diffraction amplitude can be written as a product of two functions of the three variables; the functions correspond to the sinc function and Dirichlet kernel used in the basic theory of diffraction gratings. The new expression makes it clear that hexagonally segmented telescopes are equivalent to diffraction gratings in terms of mathematical formulae. 
\end{abstract}


%%%%%%%%%%%%%%%%%%%%%%%%%%%%%%%%%%%%%%%%%%%%%%%%%%
\section{Introduction}
\label{intro}
Large telescopes have two merits: they send more light to detectors and enhance spatial resolution. However, building large telescopes using just one mirror is technically difficult, so segmented mirrors are useful for building large telescopes. The largest ground-based, single-mirror telescope is the Large Binocular Telescope (LBT) \citep{hill2006large} with an 8.4 m aperture diameter. All larger-diameter ground-based telescopes are segmented \citep{buckley2001salt,hill2004performance,geyl2004gran,mast1988keck}. In addition, future telescopes, such as the James Webb Space Telescope (JWST) \citep{codona2015james}, Thirty Meter Telescope (TMT) \citep{nelson2008status}, European Extremely Large Telescope (E-ELT) \citep{comley2011grinding}, and Giant Magellan Telescope (GMT) \citep{johns2012giant}, are also segmented.

T. S. Mast and J. E. Nelson proposed using the hexagonally packed segmented mirror to construct the Ten Meter Telescope (the current Keck Telescopes) in the late 1970s \citep{mast1979figure}.
There are some numerical calculations for evaluating its Point-Spread Function (PSF): the squared modulus of the Fourier transform of pupil functions \citep[e.g.][]{yaitskova2002tip,Neyman07kecktelescope,codona2015james}. There are also the several analytical approaches  \citep[e.g.][]{nelson1985design,zeiders1998diffraction,chanan1999strehl,yaitskova2003analytical}. Accurate numerical calculations require significant amounts of time and memory capacity; thus, the Fast Fourier Transform (FFT), an optimized algorithm of Discrete Fourier Transform (DFT), is not always the best solution. \citep{dong2013diffractive,label1529}.   

\begin{figure}
\begin{center}
\includegraphics[width=80mm]{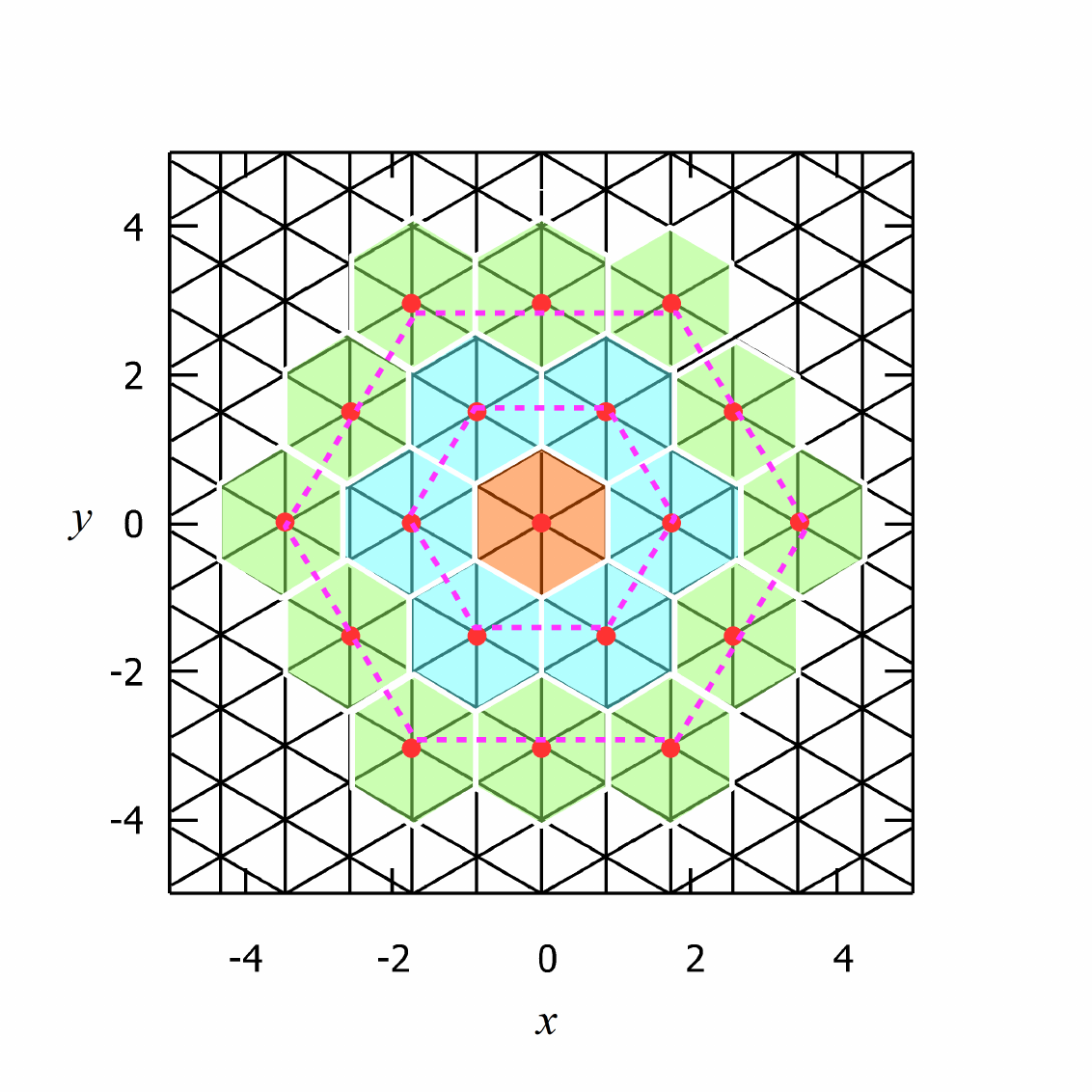}
\caption{Explanation of a hexagonally-truncated triangular grid function. A hexagonally-truncated triangular grid function is a sum of the delta functions whose peaks are located on the pupil plane. In this figure, the red dots located on segment centre indicate the delta function peak. In the case of $N=2$, the telescope consists of the green, cyan, and orange hexagonal segments. In the case of $N=1$, the telescope consists of the cyan and orange hexagonal segments. In the case of $N=0$, the telescope consists of the orange hexagonal segment.  Generally, the positions of the delta functions' peaks are all the grid-points encircled by a hexagon. This envelop hexagon is obtained by rotating the unit hexagon by $\frac{\pi}{2}$ radians and enlarging it to have the side length of $N\sqrt{3}$ (e.g. the magenta dashed hexagons (N=1,2)). Using the three variables defined in Section \ref{3var}, the area encircled by the enveloping hexagon is expressed by $|A|,|B|,|C|\leq N\sqrt{3}$. Thus, the grid-points included in this area are expressed by $A=\sqrt{3}K,\ B=\sqrt{3}L,$ and $C=\sqrt{3}M$, where $K$, $L$, $M$ are integers;  $K+L+M=0$; $|K|,|L|,|M|\leq N$. \label{abcgrid}}
\end{center}
\end{figure}

The pupil function (or aperture function) for the hexagonally segmented telescope is basically a convolution of the pupil function for a regular hexagonal aperture and a hexagonally-truncated triangular grid function (Figure \ref{abcgrid}). Thus, due to the convolution theorem, the Fourier transform of the pupil functions of the hexagonally segmented telescopes basically becomes a product of the Fourier transform of the two functions \citep{nelson1985design,zeiders1998diffraction,chanan1999strehl,yaitskova2003analytical}. This calculation is equivalent to the basic theory of diffraction gratings \citep[e.g.][]{born2000principles}, where the hexagonally segmented telescopes are equivalent to diffraction gratings \citep{yaitskova2003diffraction}.

Studies of the Fraunhofer diffraction of regular polygons include an analytical expression in the polar coordinate system \citep{komrska1972fraunhofer}, analytical expressions as a superposition of arbitrary isosceles triangles, trapezoids \citep{smith1974diffraction}, and  arbitrary triangles \citep{Sillitto:79}, and an analytical expression  with Abbe transform \citep{komrska1982simple}.  
 
On the other hand, \citet{zeiders1998diffraction} derived an analytic solution for the Fourier transform of the hexagonally-truncated triangular grid function, but do not explain how.

In this paper, the PSF of hexagonally segmented telescopes is newly formulated.
In Section 2, a key equation for the formulation is derived. In Section 3, the PSF is derived and some interpretations of the resultant equation are discussed. 

\section{Fourier Transform with Regular-triangular Symmetry} 
\label{}
\subsection{Underlying Mechanism}
\begin{figure}
\begin{center}
\includegraphics[width=30mm]{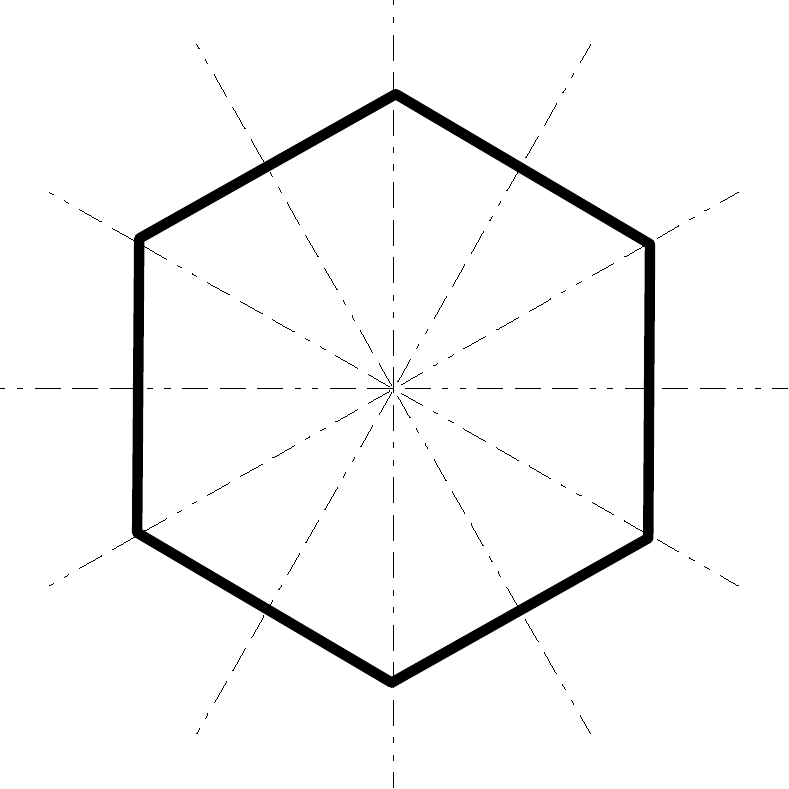}
\includegraphics[angle=90 ,width=30mm]{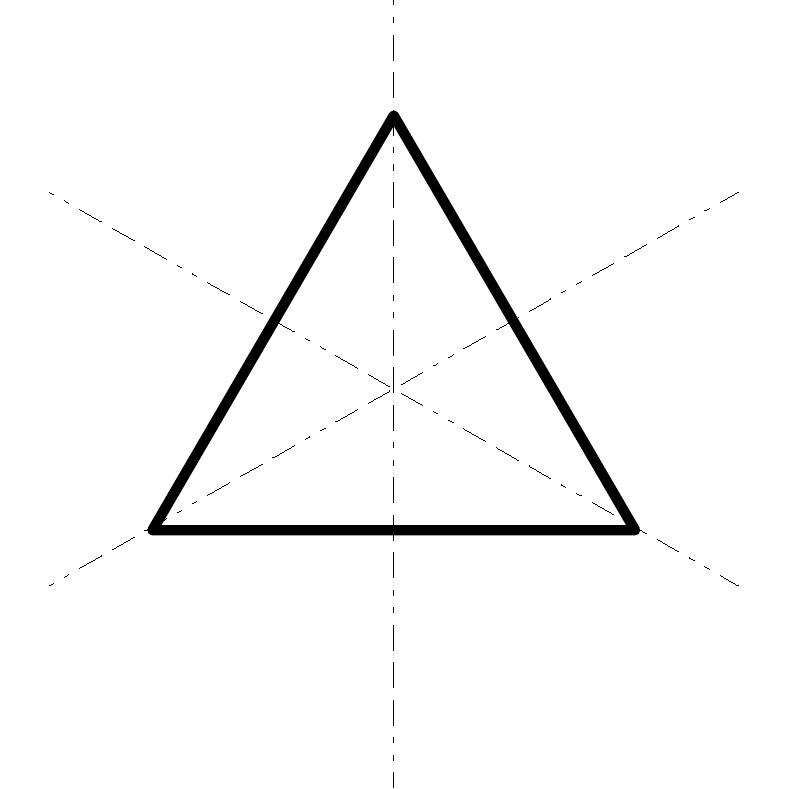}
\caption{Reflection symmetry of a regular hexagon and regular triangle. The dashed lines indicate the axes of the reflection transforms included in the symmetry operation. \label{hMirror}}
\end{center}
\end{figure}

In this study, the Fraunhofer diffraction of the hexagonally segmented telescopes is used to derive the PSF. The underlying mechanism of this work involves using a regular triangle's symmetry with the three variables on the pupil plane.
This is because, as can be seen below, the pupil function with regular-hexagonal symmetry also has regular-triangular symmetry.

\begin{table}
	\centering
	\caption{\label{RS}Symmetric operations of a regular hexagon (left-hand side) expressed by the synthesis operation of symmetric operations of a regular triangle and rotation by $\pi$ radians (right-hand side).}
	\label{operation}
   		\begin{tabular}{r|l}
        $R(\theta)$: &rotation by $\theta$  radians\\
        $S(\theta)$: &reflection for the line expressed by $y=\tan(\theta)x$  \\
        $\circ$: &symbol that represents the synthesis of operations  \\
        \end{tabular}
	\scalebox{0.85}{
    \begin{tabular}{ccc} % four columns, alignment for each
		\hline\hline
       $R\left(0\right)$&$=$&$R\left(0\right)$\\\hline
       $R\left(\frac{\pi}{3}\right)$&$=$&$R\left(\pi\right)\circ R\left(\frac{4\pi}{3}\right)$\\	\hline
       $R\left(\frac{2\pi}{3}\right)$&$=$&$R\left(\frac{2\pi}{3}\right)$\\ \hline
       $R\left(\frac{3\pi}{3}\right)$&$=$&$R\left(\pi\right)$\\ \hline
       $R\left(\frac{4\pi}{3}\right)$&$=$&$R\left(\frac{4\pi}{3}\right)$\\ \hline
       $R\left(\frac{5\pi}{3}\right)$&$=$&$R\left(\pi\right)\circ R\left(\frac{2\pi}{3}\right)$\\ \hline\hline
       \end{tabular}
       }
       \scalebox{0.85}{
   	\begin{tabular}{ccc} % four columns, alignment for each
    \hline\hline
       $S\left(0\right)$&$=$&$S\left(0\right)$\\    \hline
       $S\left(\frac{\pi}{6}\right)$&$=$&$R\left(\pi\right) \circ S\left(\frac{2\pi}{3}\right)$\\      \hline
       $S\left(\frac{2\pi}{6}\right)$&$=$&$S\left(\frac{\pi}{3}\right)$\\   \hline    
       $S\left(\frac{3\pi}{6}\right)$&$=$&$R\left(\pi\right) \circ S\left(0\right)$\\  \hline     
       $S\left(\frac{4\pi}{6}\right)$&$=$&$S\left(\frac{2\pi}{3}\right)$\\ \hline
       $S\left(\frac{5\pi}{6}\right)$&$=$&$R\left(\pi\right) \circ S\left(\frac{\pi}{3}\right)$\\ \hline
       \hline
       	\end{tabular}
        }
    	\end{table}
 
According to H. Weyl, when a figure's shape and position are maintained even if the figure is converted by an operation, it can be said that the figure is symmetric concerning the operation \citep{weyl1952symmetry}. 
These operations are referred to as 'symmetry operation' throughout this paper. 
The symmetry operations of a regular hexagon are the rotations by  $\frac{\pi n}{3}\ (n=0,1,2,3,4,5)$ radians and the reflections for the straight line expressed by $y=\tan\left(\frac{\pi n}{6}\right)x\ (n=0,1,2,3,4,5)$ (Figure \ref{hMirror}). 
Every symmetry operation of a regular hexagon can be represented by synthesizing the symmetry operations of a regular triangle and $\pi$ radian rotation (Table \ref{RS}). The symmetry operations of a regular triangle are the rotations by  $\frac{2\pi n}{3}\ (n=0,1,2)$ and the reflections for the straight line of $y=\tan\left(\frac{\pi n}{3}\right)x\ (n=0,1,2)$ (Figure \ref{hMirror}). Actually, the symmetry operations of a regular triangle are permutations of its vertices. The permutations of the regular triangle vertices have one-to-one correspondence with the permutations of the variables of functions in calculations. This one-to-one correspondence is realized by using not two, but three, variables (see Appendix \ref{impossibility}). When there is this one-to-one correspondence, functional symmetry in resultant expressions is formed such that they reflect the symmetry of the pupil plane.  Symmetry of function, as considered here, is a property that allows the values of a function to be unchanged even if its variables are permuted. For example, when a function, $f(x,y)$, satisfies the equation, $f(x,y)=f(y,x)$, it is symmetric for the permutation of two variables.

\subsection{Definition}
\label{prep}
Hereafter, normalized Cartesian coordinate systems are used both on the pupil and image planes. The coordinates are represented by $(x,y)$ on a pupil plane and $(\alpha,\beta)$ on an image plane. Both planes are perpendicular to the optical axis. The $x$-axis and  $\alpha$-axis are parallel to each other, and as are the $y$-axis and $\beta$-axis. $(x,y)$ and $(\alpha,\beta)$ are normalized by $\frac{\lambda f}{\Lambda}$ and $\Lambda$, respectively. $\lambda$, $f$, and $\Lambda$ are the wavelength of light, focal length, and side-length of the unit hexagonal segment, respectively.

Then, the delta function (see Appendix \ref{delta}) is defined as follows:
\begin{equation}
\delta (x) = \lim_{T\to \infty}\frac{\sin(\pi T x)}{\pi x}.
\end{equation}
 
In addition, a complex rectangular function is defined as follows:
\begin{equation}
\mathrm{Rect}(z) = \lim_{U \to \infty}\Pi(z)\ \ \mathrm{(}U\ \mathrm{is \ a\ natural\ number)}
\end{equation}
\begin{equation}
\Pi(z) = \frac{1}{(2z)^{2U}+1}.
\end{equation}
When $z$ is a real number denoted by $x$, $\mathrm{Rect}(x)$ is the usual rectangular function as follows:
\begin{eqnarray}
\mathrm{Rect}(x)=\left\{
\begin{array}{ll}
\frac{1}{2} & \left(|x|=\frac{1}{2} \right)\\
0 & \left(|x|>\frac{1}{2} \right)\\
1 & \left(|x|<\frac{1}{2} \right)
\end{array}
\right. .
\end{eqnarray}

\subsection{Selection of Three Variables}
\label{3var}
\begin{figure}
\begin{center}
\subfigure[]{
\includegraphics[width=60mm]{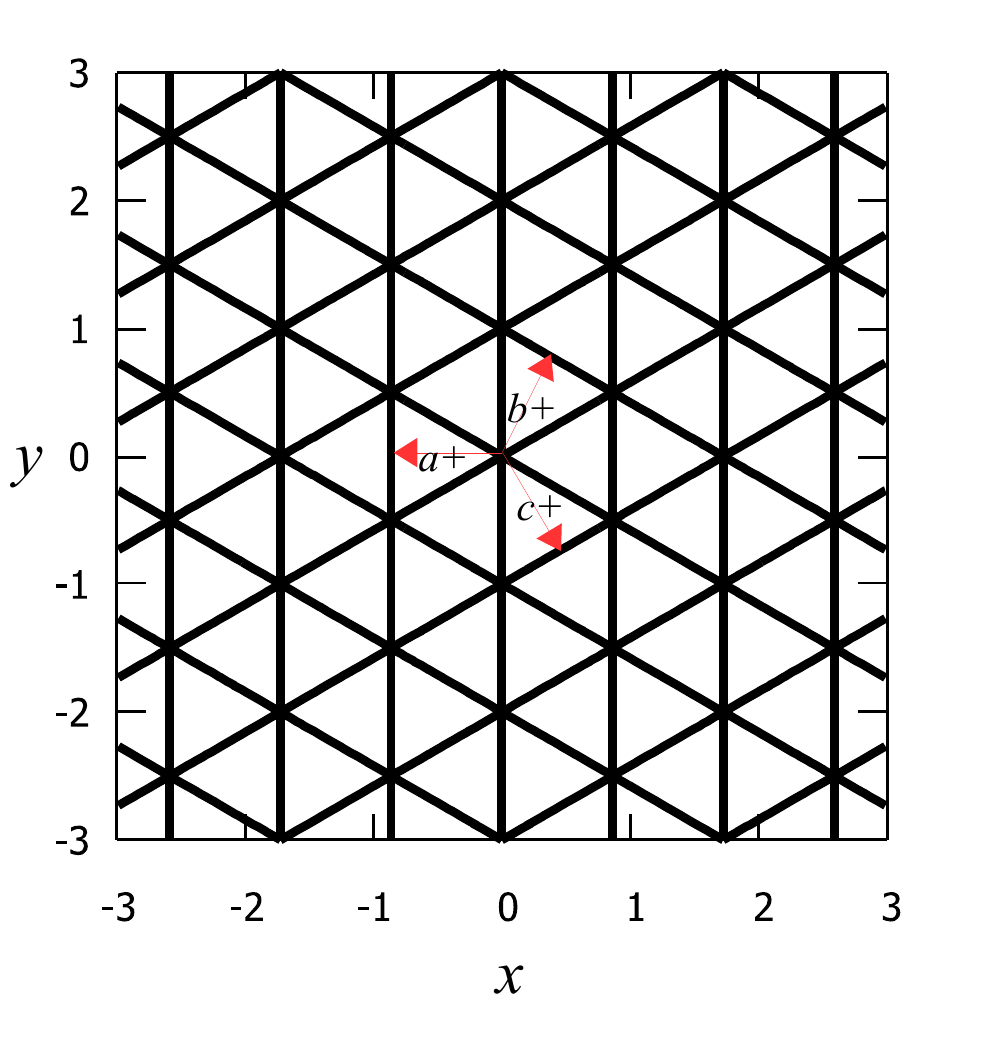}
}
\subfigure[]{
\includegraphics[width=60mm]{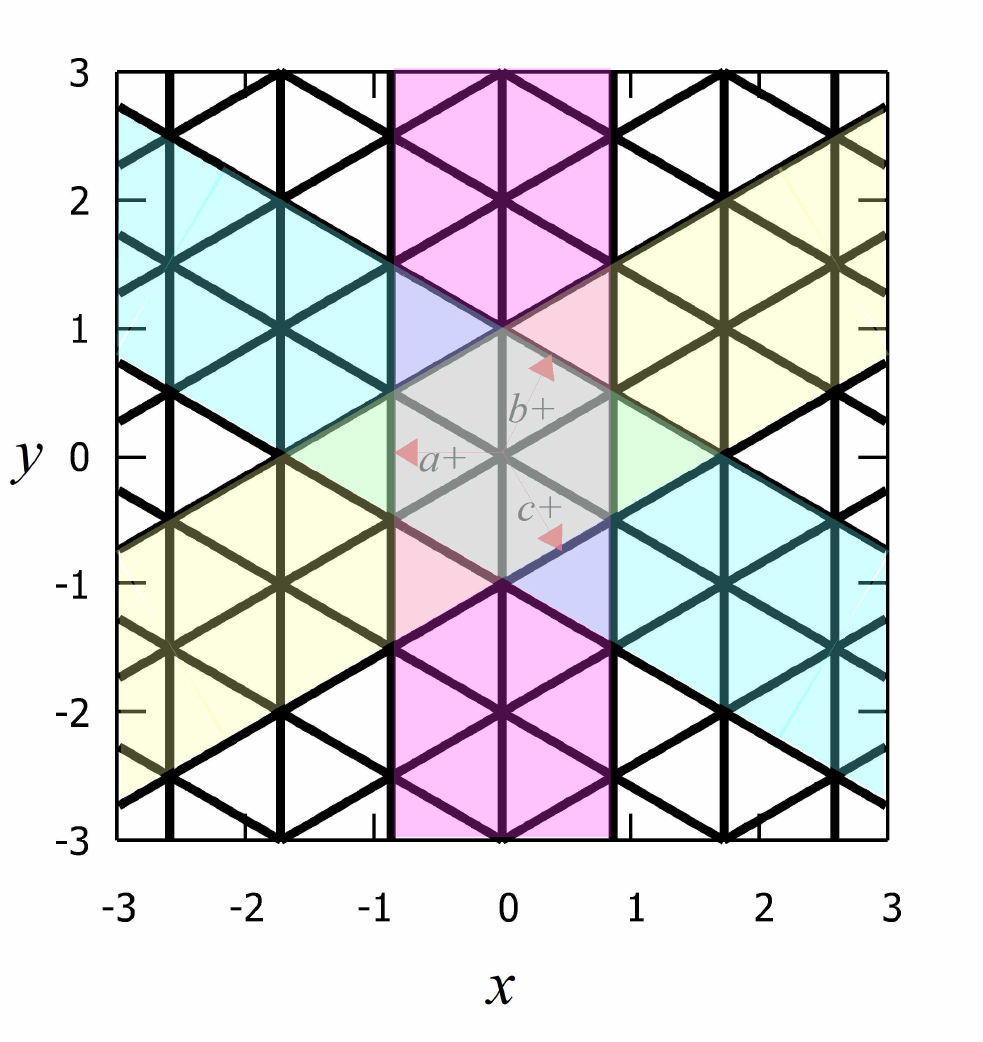}
}
\end{center}
\caption{(a) Meaning of the variables, $a$, $b$, and $c$. The black lines indicate the contour of  $a=i$, $b=j$, and $c=k$ ($i$, $j$, and $k$ are integers). The red arrows labeled by $a+$, $b+$, and $c+$ indicate the direction in which $a$, $b$, and $c$  get larger, respectively. (b) Explanation of the pupil function of a unit hexagon, as expressed by Equation (\ref{Pu1}). The semi-transparent magenta, cyan, and yellow regions indicate the regions that satisfy $|a|\leq 1 $, $|b|\leq 1 $, and $|c|\leq 1 $, respectively. The unit regular hexagonal aperture is the area in which they overlap, shown in light gray. 
\label{PupilSingle}
\label{abc}
}
\end{figure}

Three variables on the pupil plane, $a$, $b$, and $c$,  are selected as follows (Figure \ref{abc}):
\begin{eqnarray}
a&=&-\frac{2}{\sqrt{3}}x \nonumber \\
b&=&\frac{1}{\sqrt{3}}x+y \nonumber \\
c&=&\frac{1}{\sqrt{3}}x-y .
\label{defabc}
\end{eqnarray}
In the same manner, in order to satisfy Equation (\ref{inpro}), three variables on the image plane, $p$, $q$, and $r$,  are also selected as shown below:
\begin{eqnarray}
q&=&-\frac{1}{\sqrt{3}}\alpha \nonumber \\
r&=&\frac{1}{2\sqrt{3}}\alpha+\frac{1}{2}\beta \nonumber \\
s&=&\frac{1}{2\sqrt{3}}\alpha-\frac{1}{2}\beta .
\label{qrs}
\end{eqnarray}
In addition, replacing the lower-case letter, e.g. $a$, with its upper-case counterpart, e.g.  $A$, denotes a rotation by $\frac{\pi}{2}$ radians. This operation swaps the direction of the vertex and the centre of the regular hexagonal side: 
\begin{eqnarray}
A&=&-\frac{2}{\sqrt{3}}y \nonumber \\
B&=&-\frac{1}{\sqrt{3}}y+x \nonumber \\
C&=&-\frac{1}{\sqrt{3}}y-x 
\end{eqnarray}
\begin{eqnarray}
Q&=&\frac{1}{\sqrt{3}}\beta \nonumber \\
R&=&-\frac{1}{2\sqrt{3}}\beta+\frac{1}{2}\alpha \nonumber \\
S&=&-\frac{1}{2\sqrt{3}}\beta-\frac{1}{2}\alpha .
\label{QRS}
\end{eqnarray}
These variables satisfy the following equation:
\begin{equation}
a+b+c=A+B+C=q+r+s=Q+R+S=0 .
\label{sumzero}
\end{equation}
 
The equations for the definitions of $a$, $b$, and $c$, can be interpreted as contour line equations of $a$, $b$, and $c$, respectively. Those straight lines are parallel to each side of a regular triangle; the same is true for the other sets of variables defined above. 

There is a one-to-one correspondence between the permutations of these variables and the symmetry operations of a regular triangle. For example, transposing $a$ and $b$ $\left[(a,b,c) \to (b,a,c)\right]$ indicates a reflection transform. Cyclic permutation, that is $(a,b,c)\to (c,a,b)$, indicates the rotation by $\frac{2\pi}{3}$ radians. On the other hand, sign inversion $(a,b,c) \to (-a,-b,-c)$ indicates the rotation by $\pi$ radians. Hence, every symmetric operation of a regular hexagon is expressed by the synthesis of the sign inversion and permutations of the three variables. 

The diffraction amplitude for the hexagonally segmented telescope is a product of the diffraction amplitude of a single unit regular hexagonal aperture and a hexagonally-truncated triangular grid function \citep{nelson1985design,zeiders1998diffraction,chanan1999strehl,yaitskova2003analytical}.
The pupil function of a unit hexagon (Figure \ref{PupilSingle}) is 
\begin{equation}
\bar{P}_{u}(a,b,c)=\mathrm{Rect}\left(\frac{a}{2}\right)\mathrm{Rect}\left(\frac{b}{2}\right)\mathrm{Rect}\left(\frac{c}{2}\right).
\label{Pu1}
\end{equation}

The pupil function of the hexagonally-truncated hexagonal gird function (Figure \ref{abcgrid})  is
\begin{equation}
\bar{P}_{g}(a,b,c)=\sum_{|K|,|L|,|M|\leq N}\delta(A-\sqrt{3}K)\delta(B-\sqrt{3}L)u(C-\sqrt{3}M),
\label{Pu2}
\end{equation}
where $K+L+M=0$ and $u(x)= \lim_{T\to \infty}\frac{\sin(\pi T x)}{\pi T x}. $

\begin{figure}
\begin{center}
\includegraphics[width=80mm]{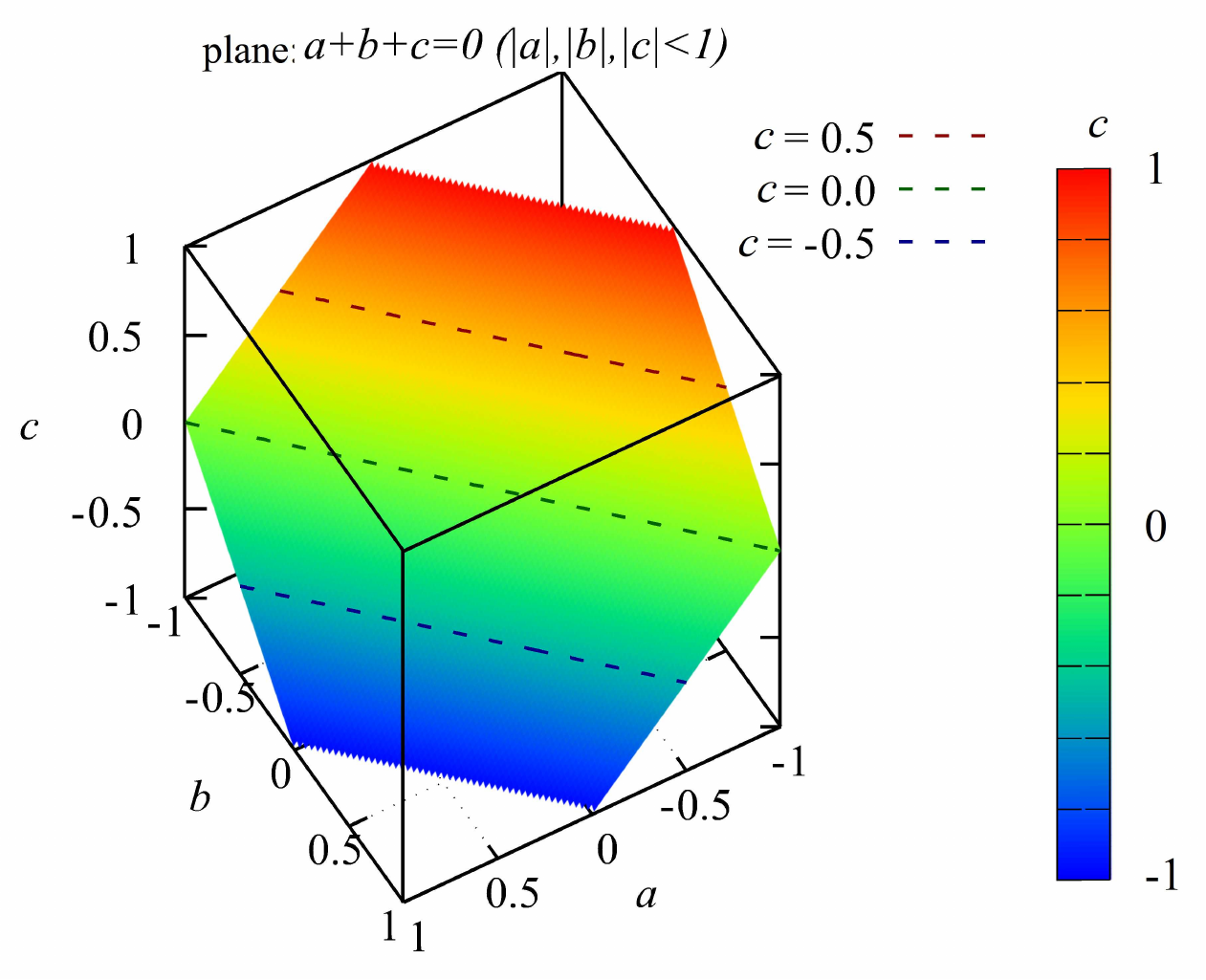}
\caption{By considering the three variables defined by Equation (\ref{defabc}) as the coordinates of the 3-D Cartesian coordinate system, Equation (\ref{Pu1}) represents a regular hexagon on the plane expressed by Equation (\ref{sumzero}); the color map and contour lines indicate the value of $c$ on the regular hexagon. When the condition of Equation (\ref{sumzero}) is ignored, Equation (\ref{Pu1}) represents a cube (regular hexahedron), and the black solid frame indicates the sides of the cube.  \label{abc3d}}
\end{center}
\end{figure}

When $(a,b,c)$ are interpreted as coordinates of the 3-D Cartesian system (see Figure \ref{abc3d}), while ignoring the condition of Equation (\ref{sumzero}), the value of Equation (\ref{Pu1}) becomes unity inside a cube and zero outside the cube. 
In the same manner, Equation (\ref{Pu2}) becomes a sum of delta functions whose peaks are located on the cubically-truncated 3-D grid. These functions of the three variables freed from the condition of Equation (\ref{sumzero}), are called cubic 'pupil functions' throughout this paper.

This equation represents the inner product of $(\alpha,\beta)$ and $(x,y)$ with the variables defined above:
\begin{equation}
qa+rb+sc=QA+RB+SC=\alpha x+\beta y.
\label{inpro}
\end{equation}
The following equations are also used to relate the variables denoted by lower-case and upper-case letters with each other:
\begin{equation}
A=\frac{b-c}{\sqrt{3}},\ \ B=\frac{c-a}{\sqrt{3}},\ \ C=\frac{a-b}{\sqrt{3}} 
\end{equation}
\begin{equation}
Q=\frac{r-s}{\sqrt{3}},\ \ R=\frac{s-q}{\sqrt{3}},\ \ S=\frac{q-r}{\sqrt{3}} 
\label{LS}
\end{equation}
\begin{equation}
a=\frac{B-C}{-\sqrt{3}},\ \ b=\frac{C-A}{-\sqrt{3}},\ \ c=\frac{A-B}{-\sqrt{3}} 
\end{equation}
\begin{equation}
q=\frac{R-S}{-\sqrt{3}},\ \ r=\frac{S-Q}{-\sqrt{3}},\ \ s=\frac{Q-
R}{-\sqrt{3}} .
\end{equation} 
\subsection{Formulation of Fourier Transform with Regular-triangular Symmetry}
\label{FF}
The Fourier transform of a pupil function into the 3-D Fourier transform of a cubic pupil function is now developed.
Here, $P(x,y)=\bar{P}(a(x,y),b(x,y),c(x,y))$ denotes a pupil function, and $v(\alpha,\beta)$ denotes the Fourier transform of the pupil function. By using Equations (\ref{inpro}) and (\ref{sumzero}), the following equation is obtained:
\begin{eqnarray}
&&v(\alpha,\beta) = \int \!\!\! \int _{-\infty}^{\infty}\!\!\!  dx dy P(x,y)e^{-2\pi i (\alpha x+\beta y)} \nonumber \\
&=&\!\!\!\!\!\frac{\sqrt{3}}{2}\int \!\!\!\int _{-\infty}^{\infty}\!\!\!  da db \bar{P}(a,b,-a-b)e^{-2\pi i (qa+rb+s(-a-b))}.
\label{ft}
\end{eqnarray}

By using delta function, Equation (\ref{ft}) becomes
\begin{eqnarray}
v(\alpha,\beta)\!\!\!\!\!\!&=&\!\!\!\!\!\!\frac{\sqrt{3}}{2}\int \!\!\! \int \!\!\! \int _{-\infty}^{\infty}\!\!\!  da db dc  \bar{P}(a,b,c) \delta(a+b+c) e^{-2\pi i (qa+rb+sc)}.\nonumber \\ 
\end{eqnarray}
Here, the integral variables,  $a$, $b$, and $c$, are no longer limited to the region , $a+b+c=0$.

Since the inner products of the two functions are equal to the inner products of the Fourier transforms of the two functions,  $v(\alpha,\beta)$ becomes
\begin{eqnarray}
v(\alpha,\beta)\!\!\!\!\!\!&=&\!\!\!\!\!\!\frac{\sqrt{3}}{2}\int \!\!\! \int \!\!\! \int _{-\infty}^{\infty}\!\!\!  dq' dr' ds'\left(\int \!\!\! \int _{-\infty}^{\infty}\!\!\!  da db e^{-2\pi i ((q'-s')a+(r'-s')b)}\right)^{\ast}\nonumber \\
&\times&\!\!\!\!\!\! \left(\int \!\!\! \int \!\!\! \int _{-\infty}^{\infty}\!\!\!  da db dc  \bar{P}(a,b,c)  e^{-2\pi i ((q'+q)a+(r'+r)b+(s'+s)c)}\right). \nonumber \\
\end{eqnarray}
Hence,
\begin{eqnarray}
v(\alpha,\beta)\!\!\!\!\!\!&=&\!\!\!\!\!\!\frac{\sqrt{3}}{2}\int \!\!\! \int \!\!\! \int _{-\infty}^{\infty}\!\!\!  dq' dr' ds' \delta(q'\!-\!s')\delta(r'\!-\!s') \nonumber \\
&\times& \left(\int \!\!\! \int \!\!\! \int _{-\infty}^{\infty}\!\!\!  da db dc  \bar{P}(a,b,c)  e^{-2\pi i ((q'+q)a+(r'+r)b+(s'+s)c)}\right) \nonumber \\
\!\!\!\!\!\!\!\!\!\!\!\!=\!\!\!\!\!\!\!\!\!\!\!&&\!\!\!\!\!\!\!\frac{\sqrt{3}}{2}\int _{-\infty}^{\infty}\!\!\!\!\!\!  d\tau \!\!\!\int \!\!\!\!\! \int \!\!\!\!\! \int _{-\infty}^{\infty}\!\!\!\!\!  da db dc  \bar{P}(a,b,c)  e^{-2\pi i ((q+\tau)a+(r+\tau)b+(s+\tau)c)},\nonumber\\
\label{BASICbefore}
\end{eqnarray}
where $\tau=s'$ has been redefined.

When the variables $(a,b,c)$ are interpreted as coordinates of a 3-D Cartesian coordinate system, 
\begin{equation}
\int \!\!\! \int \!\!\! \int _{-\infty}^{\infty}\!\!\!  da db dc  \bar{P}(a,b,c)  e^{-2\pi i ((q+\tau)a+(r+\tau)b+(s+\tau)c)}
\end{equation}
is a 3-D Fourier transform of $\bar{P}(a,b,c)$. In the 3-D Cartesian coordinate system, this integral, $\int _{-\infty}^{\infty}\!\!\!  d\tau$, indicates that this line integral along a straight line is passing through the point, $(a,b,-a-b)$ and is perpendicular to the plane $a+b+c=0$.

The right-hand sides of Equations (\ref{Pu1}) and (\ref{Pu2}) can be written as $f(a)f(b)f(c)$ by a function denoted by $f(a)$.
When $\bar{P}(a,b,c)$ can be written as $f(a)f(b)f(c)$, the 3-D Fourier transform of $\bar{P}(a,b,c)$ becomes $\tilde{f}(q)\tilde{f}(r)\tilde{f}(s)$, where $\tilde{f}(q)$ denotes the Fourier transform of $f(x)$. Thus, Equation (\ref{BASICbefore}) becomes
\begin{eqnarray}
v(\alpha,\beta)&=&\frac{\sqrt{3}}{2}
\int _{-\infty}^{\infty}\!\!\!  d\tau \prod_{C\left\lbrace q,r,s\right\rbrace} \tilde{f}(q+\tau), 
\label{BASIC}
\end{eqnarray}
where $\prod_{C\left\lbrace a,b,c\right\rbrace}$ indicates the product over the cyclic permutations of $\left\lbrace a,b,c\right\rbrace$.
\begin{figure}
\begin{center}
\includegraphics[width=60mm]{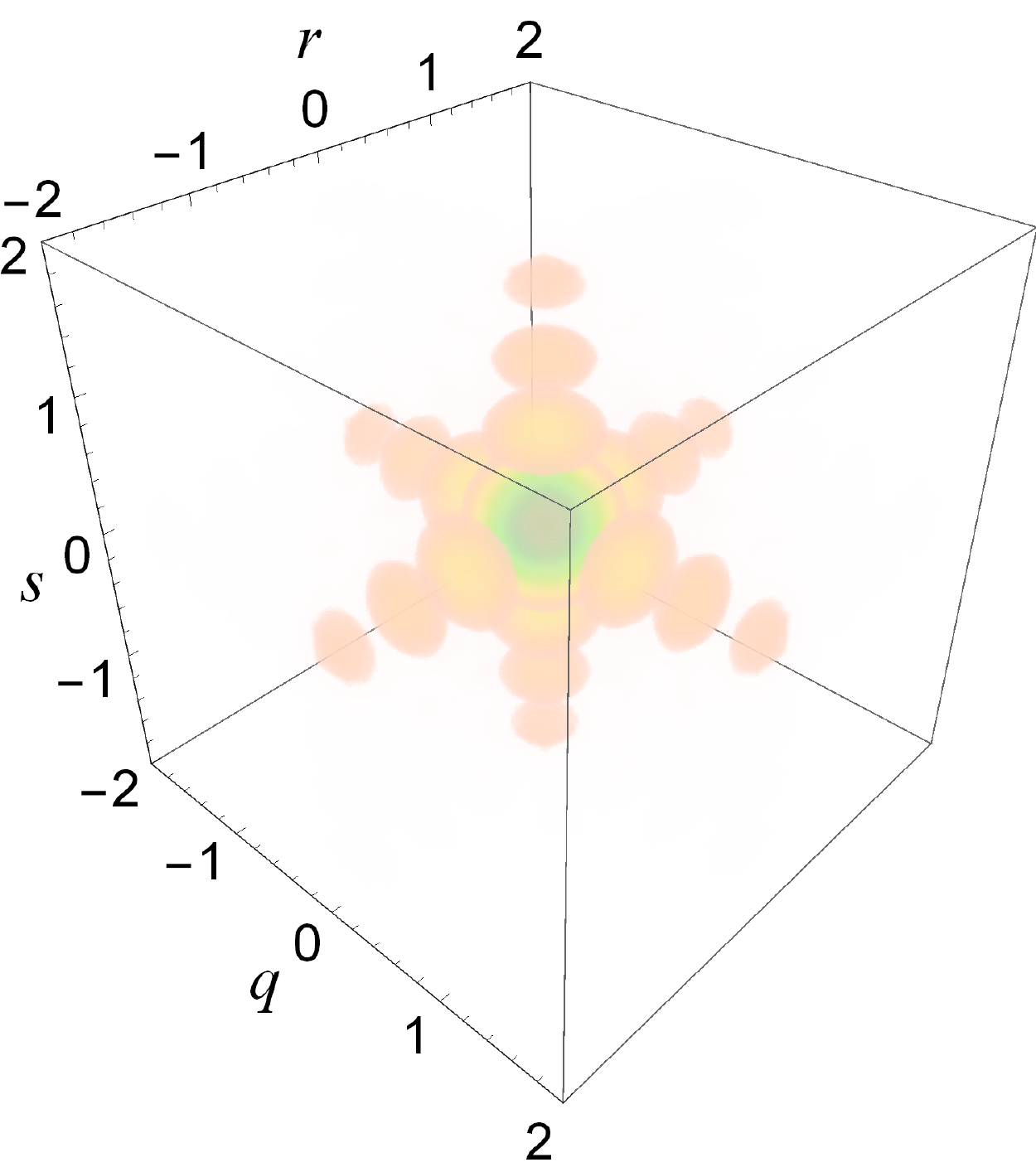}
\includegraphics[width=12mm]{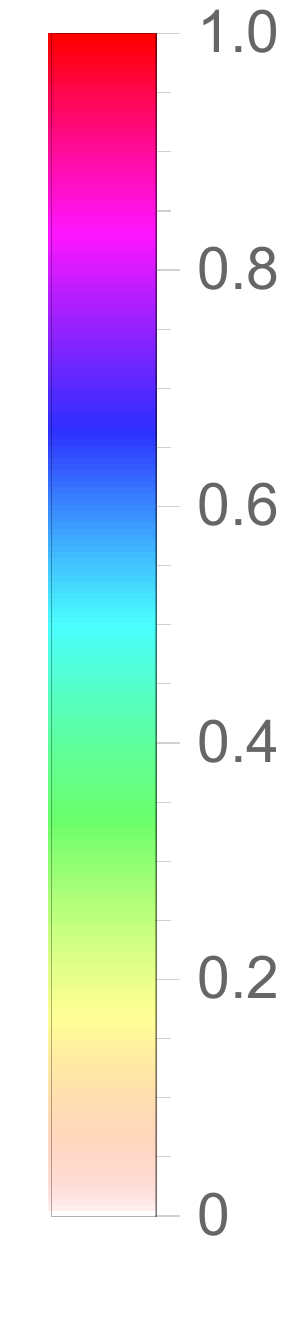}
\caption{3-D plot of $\frac{\sin(2\pi q)}{\pi q}$$\frac{\sin(2\pi r)}{\pi r}$$\frac{\sin(2\pi s)}{\pi s}$. This function is normalized by the value of the origin and is shown in absolute values. The line integral along a line perpendicular to the plane, $q+r+s=0$, gives an expression for the diffraction amplitude of a unit regular-hexagonal aperture.  \label{qrssinc}}
\end{center}
\end{figure}
  
For example, $\tilde{f}(q)$ basically become the form of sinc function (Figure \ref{qrssinc}), $\frac{\sin x}{x}$ \citep{howell2001principles}, and Dirichlet kernel, $\frac{\sin\left(\left(n+\frac{1}{2}\right) x\right)}{\sin \left(\frac{x}{2}\right)}$ \citep{howell2001principles}, for the pupil functions of Equations (\ref{Pu1}) and (\ref{Pu2}), respectively. Both of these functions are used in basic diffraction grating theory \citep[e.g.][]{born2000principles}. 

The same equation is true for the Fourier transform of $\bar{P}(A,B,C)$, if $(q,r,s)$ in Equation (\ref{BASIC}) is replaced by $(Q,R,S)$.
\section{Formulation of PSF of Hexagonally Segmented Telescope}
In this section, we derive an analytical expression for the diffraction amplitude of hexagonally segmented telescopes. The diffraction amplitude, $v_c(\alpha,\beta)$, can be basically written as the product of the Fourier transform of a unit regular-hexagonal aperture, $v_u(\alpha,\beta)$, and a hexagonally-truncated triangular grid function (Figure \ref{abcgrid}), $v_g(\alpha,\beta)$, as follows \citep[e.g.][]{nelson1985design}:
\begin{equation}
v_c(\alpha,\beta)=\mathrm{Scale}\left[v_u(\alpha,\beta)\right]v_g(\alpha,\beta),
\label{FP1}
\end{equation}
where the scaling transformation, $\mathrm{Scale}\left[v_u(\alpha,\beta)\right]$, is defined as 
\begin{equation}
\mathrm{Scale}\left[v_u(\alpha,\beta)\right]=(1-\Delta)^2v_u\left((1-\Delta)\alpha,(1-\Delta)\beta \right),
\label{FP2}
\end{equation}
and $\Delta$ denotes $\frac{2}{\sqrt{3}}$ times the half width of the gap between adjacent segments; $1-\Delta$ corresponds to the side-length of a segment. 
\subsection{Unit Regular Hexagon}
\label{one}
First, Equation (\ref{BASIC}) is applied to the pupil function of regular-hexagonal aperture whose side length is unity. The pupil function, $\bar{P_u}(a,b,c)$, is shown in Equation (\ref{Pu1}); the Fourier transform of the rectangular function becomes the sinc function as follows: 
\begin{eqnarray}
\tilde{f_u}(q)&=&\int^{\infty}_{-\infty}da\ \mathrm{Rect}\left(\frac{a}{2}\right)e^{-2\pi i q a}\nonumber\\&=&\frac{\sin(2\pi q)}{\pi q}.
\label{FourSinc}
\end{eqnarray}
Hence, Equation (\ref{BASIC}) becomes
\begin{eqnarray}
v_u(\alpha,\beta)&=&
\frac{\sqrt{3}}{2}\int _{-\infty}^{\infty}\!\!\!  d\tau \prod_{C\left\lbrace q,r,s\right\rbrace}\tilde{f_u}(q+\tau)\nonumber \\
\begin{comment}
&=&
\frac{\sqrt{3}}{2}\int _{-\infty}^{\infty}\!\!\!  d\tau \prod_{C\left\lbrace q,r,s\right\rbrace}\frac{+e^{+2\pi(q+\tau)}-e^{-2\pi(q+\tau)}}{(2i)\pi(q+\tau)}\nonumber \\ 
\end{comment}
&=&\frac{\gamma}{2\pi i} \sum_{\sigma_1=\pm 1}\sum_{\sigma_2=\pm 1}\sum_{\sigma_3=\pm 1}\sigma_1 \sigma_2 \sigma_3 e^{2\pi i (\sigma_1 q+\sigma_2 r+\sigma_3 s)}\nonumber \\
&\times& \int _{-\infty}^{\infty}\!\!\!  dz g(z), 
\label{single}
\end{eqnarray}
where $\tau \in \mathbb{R}$ has been substituted with $z\in \mathbb{C}$ , and $\gamma$ and $g(z)$ are defined as follows: 
\begin{equation}
\gamma = \frac{\sqrt{3}}{2}(2\pi i)^{-2}
\label{gamma}
\end{equation}
\begin{equation}
g(z) = \frac{e^{2\pi i(\sigma_1 +\sigma_2 +\sigma_3)z}}{(q+z)(r+z)(s+z)}.
\end{equation}
 
The right-hand side of Equation (\ref{single}) can be evaluated for all $(q,r,s)$ by dividing into three cases as follows; (i) $q-r$, $r-s$, and $s-q$ are all zero, (i\hspace{-.1em}i) one of $\left\lbrace q-r\right.$, $r-s$, $\left.s-q \right\rbrace$ is zero but the others are not, and (i\hspace{-.1em}i\hspace{-.1em}i) none of $\left\lbrace q-r\right.$, $r-s$, $\left.s-q \right\rbrace$ is zero.

\begin{figure} 
\subfigure[$\sigma_1 +\sigma_2 +\sigma_3 >0$]{
\includegraphics[width=80mm,trim=6cm 16.5cm 8.5cm 5cm]{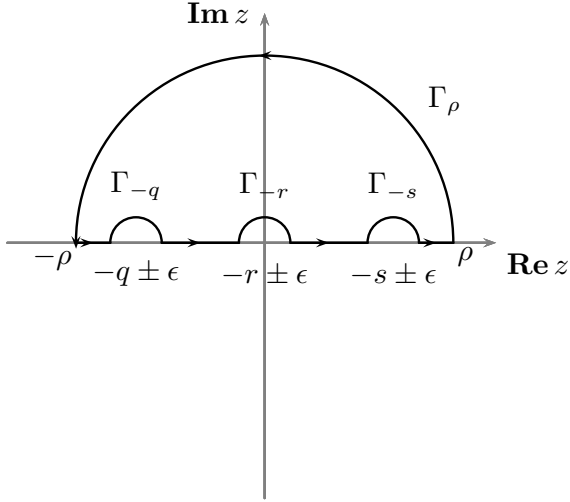}
    }  
    \subfigure[$\sigma_1 +\sigma_2 +\sigma_3 <0$]{
\includegraphics[width=80mm,trim=6cm 16.5cm 8.5cm 5cm]{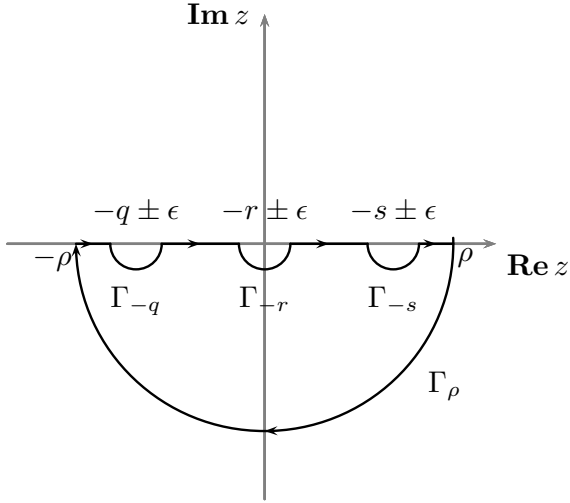}
    } 
    \caption{\label{integralpath1} (a) Integral path in case (i\hspace{-.1em}i\hspace{-.1em}i) for positive $\sigma_1$$+$$\sigma_2$$+$$\sigma_3$. The integrand must be regular inside the integral paths.  (b)Integral path in case (i\hspace{-.1em}i\hspace{-.1em}i) for negative $\sigma_1$$+$$\sigma_2$$+$$\sigma_3$. The integrand must be regular inside the integral paths. }
\end{figure}

First, case (i\hspace{-.1em}i\hspace{-.1em}i) is treated. The other cases are obtained in Appendix \ref{removable} as limiting values of the results for case (i\hspace{-.1em}i\hspace{-.1em}i).
An integral path, as shown in Figure \ref{integralpath1}, is considered for $g(z)$ in Equation (\ref{single}). There are the semicircles for avoiding singularities, denoted by $\Gamma_{-q}$, $\Gamma_{-r}$, and $\Gamma_{-s}$, and the outer semicircle denoted by $\Gamma_\rho$. The radius of $\Gamma_\rho$ approaches infinity ($\rho\to\infty$), and the radii of $\Gamma_{-q}$, $\Gamma_{-r}$, and $\Gamma_{-s}$ approach zero ($\epsilon\to0$).
By applying the Cauchy integral theorem along the integral path, the last factor of Equation (\ref{single}) can be calculated as follows:
\begin{equation}
\int _{-\infty}^{\infty}\!\!\!  dz g(z) = -\left(\int_{\Gamma_{-q}}+\int _{\Gamma_{-r}}+\int_{\Gamma_{-s}}\right) dz g(z) .
\label{gamm}
\end{equation}
When $\sigma_1 +\sigma_2 +\sigma_3$ is positive, Equation (\ref{gamm}) becomes 
\begin{eqnarray} 
&\int _{-\infty}^{\infty}&\!\!\!  dz g(z)\nonumber\\
&=& -\lim_{\epsilon \to 0} \sum_{C\left\lbrace q,r,s\right\rbrace} \int_{\pi}^0 i\epsilon e^{i\theta} d\theta g(\epsilon e^{i \theta}-q) \nonumber \\
&=& i\lim_{\epsilon \to 0} \int_{0}^{\pi}d\theta \sum_{C\left\lbrace q,r,s\right\rbrace} \frac{e^{2\pi i(\sigma_1 +\sigma_2 +\sigma_3)(-q+\epsilon e^{i \theta})}}{(r-q+\epsilon e^{i \theta})(s-q+\epsilon e^{i \theta})}\nonumber \\
\begin{comment}
&=& i \int_{0}^{\pi} d\theta \sum_{C\left\lbrace q,r,s\right\rbrace}
\frac{e^{2\pi i(\sigma_1 +\sigma_2 +\sigma_3)(-q)}}{(r-q)(s-q)} \nonumber \\
\end{comment}
&=& i\pi \sum_{C\left\lbrace q,r,s\right\rbrace}
\frac{e^{2\pi i(\sigma_1 +\sigma_2 +\sigma_3)(-q)}}{(r-q)(s-q)},
\label{plus} 
\end{eqnarray}
where $\sum_{C\left\lbrace q,r,s\right\rbrace}$ indicates the sum over the cyclic permutations of $\left\lbrace q,r,s \right\rbrace$.
When $\sigma_1 +\sigma_2 +\sigma_3$ is negative, this equation becomes
\begin{eqnarray}
&\int _{-\infty}^{\infty}&\!\!\!  dz g(z)\nonumber\\
&=& -\lim_{\epsilon \to 0} \sum_{C\left\lbrace q,r,s\right\rbrace} \int_{-\pi}^0 i\epsilon e^{i\theta} d\theta g(\epsilon e^{i \theta}-q) \nonumber \\
&=& - i\pi \sum_{C\left\lbrace q,r,s\right\rbrace}
\frac{e^{2\pi i(\sigma_1 +\sigma_2 +\sigma_3)(-q)}}{(r-q)(s-q)}.
\label{minus}
\end{eqnarray}
Equation (\ref{minus}) is equivalent to Equation (\ref{plus}), except for the sign inversion. 

In order to calculate $v_u(\alpha,\beta)$,
the following terms in Equation (\ref{single}) are summed up: 
\begin{equation}
\frac{\gamma}{2\pi i}\sigma_1 \sigma_2 \sigma_3 e^{2\pi i (\sigma_1 q+\sigma_2 r+\sigma_3 s)}\int _{-\infty}^{\infty}\!\!\!  dz g(z).
\end{equation}
\begin{table}
	\centering
	\caption{Calculation of signs contributing to the summation in Equations (\ref{single}) and (\ref{grid}).} 
	\label{sign}
    \begin{tabular}{c} % four columns, alignment for each
				 $S(\sum_{i}\sigma_i)=\left\lbrace1\ (\mathrm{for\ } \sum_{i}\sigma_i>0),\ -1\ (\mathrm{for\ }\sum_{i}\sigma_i<0)\right\rbrace$ 
			\end{tabular}
	\begin{tabular}{|l|ccc|c|c|c|c|} % four columns, alignment for each
		\hline\hline
		   &$\sigma_1$&$\sigma_2$&$\sigma_3$&$\sum_{i}\sigma_i$&$S$&$\sigma_1\sigma_2\sigma_3$&$S\sigma_1\sigma_2\sigma_3$ \\ 
		\hline
		\ctext{1} & 1 & 1 & 1 &3& 1 & 1 & 1\\
		\ctext{2} & 1 & 1 & -1 &1& 1 & -1 & -1 \\
		\ctext{3} & 1 & -1 & 1 &1& 1 & -1 & -1 \\
        \ctext{4} & 1 & -1 & -1 &-1& -1 & 1 & -1 \\
		\ctext{5} & -1 & 1 & 1 &1& 1 & -1 & -1 \\
		\ctext{6} & -1 & 1 & -1 &-1& -1 & 1 & -1 \\
        \ctext{7} & -1 & -1 & 1 &-1& -1 & 1 & -1 \\
		\ctext{8} & -1 & -1 & -1 &-3& -1 & -1 & 1 \\
		\hline\hline
	\end{tabular}
    	\end{table} 
 A part of the summation in Equation (\ref{single}) is 
\begin{eqnarray}
\mathrm{\ctext{1}+\ctext{8}}
&=&\gamma \sum_{C\left\lbrace q,r,s\right\rbrace} \frac{\cos(2\pi(r+s-2q))}{(r-q)(s-q)},
\end{eqnarray}
where the encircled numbers, such as  \ctext{1}, are identifiers of the sets indicated in Table \ref{sign}, and \ctext{1}+\ctext{8} denotes the sum of the sets, \ctext{1} and \ctext{8}.
In the same manner, the summations, \ctext{2}+\ctext{7}, \ctext{3}+\ctext{6}, and \ctext{4}+\ctext{5} are calculated as follows:
\begin{eqnarray}
&&\begin{bmatrix}
\mathrm{\ctext{2}+\ctext{7}}\\
\mathrm{\ctext{3}+\ctext{6}}\\
\mathrm{\ctext{4}+\ctext{5}}
\end{bmatrix} \nonumber\\
&=&\!\!\!\!\!\!-\gamma
\begin{bmatrix}
\frac{\cos(2\pi(r-s))}{(r-q)(s-q)}+\frac{\cos(2\pi(-s+q))}{(s-r)(q-r)}+\frac{\cos(2\pi(q+r-2s))}{(q-s)(r-s)} \\ 
\frac{\cos(2\pi(-r+s))}{(r-q)(s-q)}+\frac{\cos(2\pi(s+q-2r))}{(s-r)(q-r)}+\frac{\cos(2\pi(q-r))}{(q-s)(r-s)} \\
\frac{\cos(2\pi(-r-s+2q))}{(r-q)(s-q)}+\frac{\cos(2\pi(-s+q))}{(s-r)(q-r)}+\frac{\cos(2\pi(q-r))}{(q-s)(r-s)} 
\end{bmatrix}.
\end{eqnarray}
Thus, by substituting these sums into the right-hand side of Equation (\ref{single}),  $v_u(\alpha,\beta)$ is expressed as follows:
\begin{eqnarray}
v_u(\alpha,\beta)\!\!\!\!\!\!&=&\!\!\!\!\!\! 2\gamma \sum_{C\left\lbrace q,r,s\right\rbrace} \frac{\cos(2\pi(r-s))}{(s-q)(q-r)}.
\label{value}
\end{eqnarray}
By using Equation (\ref{LS}), Equation (\ref{value}) can be written as  
\begin{eqnarray}
v_u(\alpha,\beta)\!\!\!\!\!\!&=&\!\!\!\!\!\! 2\pi^2\gamma \sum_{C\left\lbrace Q,R,S\right\rbrace}\frac{\cos(2\pi\sqrt{3}Q)}{(\pi\sqrt{3}R)(\pi\sqrt{3}S)}.
\label{result1}
\end{eqnarray}
This is the resultant expression for the diffraction amplitude of the unit regular hexagon pupil function. 

Equation (\ref{result1}) is equivalent to some previous works concerning a unit hexagonal aperture \citep[e.g.][]{nelson1985design}. 
The proof uses the equation; $\sin(u)\sin(v)=\frac{1}{2}\left(\cos(u-v)-\cos(u+v)\right)$.
Some studies \citep{smith1974diffraction,chanan1999strehl} have obtained the analytical expression by superposing some trapezoids. It is clear that these expressions are also equivalent to Equation (\ref{result1}).

\subsection{Hexagonally-Truncated Triangular Grid Function}
\label{two}
Next, the diffraction amplitude of a hexagonally-truncated triangular grid function, with an interval of $\sqrt{3}$, is derived. The pupil function shown in Equation (\ref{Pu2}) can be written as follows:
\begin{eqnarray} 
&\bar{P_g}&\!\!\!\!\!\!(A,B,C)\nonumber \\
&=& \!\!\!\!\!\!\lim_{T \to \infty}\frac{1}{T} \prod_{C\left\lbrace A,B,C\right\rbrace}\left\lbrace\sum_{|K|\leq N}\frac{\sin(\pi T(A-\sqrt{3}K))}{\pi(A-\sqrt{3}K)}\right\rbrace.
\label{asdfg}
\end{eqnarray}
The Fourier transform of a factor in Equation (\ref{asdfg}) is a product of the rectangular function and Dirichlet kernel as follows:
\begin{eqnarray} 
\tilde{f_g}(Q)&=&\int^{\infty}_{-\infty}dA\left\lbrace\sum_{|K|\leq N}\frac{\sin(\pi T(A-\sqrt{3}K))}{\pi(A-\sqrt{3}K)}\right\rbrace e^{-2\pi i QA}\nonumber\\
&=& \mathrm{Rect}\left(\frac{Q}{T}\right)\frac{\sin((2N+1)\sqrt{3}\pi Q)}{\sin(\sqrt{3}\pi Q)}.
\end{eqnarray}
Hence, the diffraction amplitude, $\nu(\alpha, \beta)$, can be derived from Equation (\ref{BASIC}) as follows:  
\begin{eqnarray}
v_g(\alpha,\beta)&=&
\frac{\sqrt{3}}{2}\lim_{T \to \infty}\frac{1}{T}\int _{-\infty}^{\infty}\!\!\!  d\tau \prod_{C\left\lbrace Q,R,S\right\rbrace} \tilde{f_g}(Q+\tau)\nonumber \\
&=&\!\!\!\!\!\!\frac{\sqrt{3}}{2(2 i)^{3}}\sum_{\sigma_1=\pm 1}\sum_{\sigma_2=\pm 1}\sum_{\sigma_3=\pm 1}\sigma_1 \sigma_2 \sigma_3 e^{2\pi i (\sigma_1 Q+\sigma_2 R+\sigma_3 S)}\nonumber\\
&\times&\lim_{T,U \to \infty}\frac{1}{T}\int _{-\infty}^{\infty}\!\!\!  dz \Pi^3(z) h(z), 
\label{grid}
\end{eqnarray}
where $h(z)$ and $\Pi^3(z)$ are defined as follows:
\begin{equation}
h(z) = \frac{e^{(2N+1)\sqrt{3}\pi i(\sigma_1 +\sigma_2 +\sigma_3)z}}{\prod_{C\left\lbrace Q,R,S\right\rbrace}\sin(\sqrt{3}\pi(Q+z))}
\end{equation}
\begin{equation}
\Pi^3(z) = \prod_{C\left\lbrace Q,R,S\right\rbrace}\Pi\left(\frac{Q+z}{T}\right).
\end{equation}
 
The right-hand side of Equation (\ref{grid}) can be evaluated for all $(Q,R,S)$ by dividing into three cases as follows: (I) $Q-R$, $R-S$, and $S-Q$ are all $\frac{1}{\sqrt{3}}$ times integers, (I\hspace{-.1em}I) one of $\left\lbrace Q-R\right.$, $R-S$, $\left.S-Q \right\rbrace$ is $\frac{1}{\sqrt{3}}$ times integers but the others are not, and (I\hspace{-.1em}I\hspace{-.1em}I) none of $\left\lbrace Q-R\right.$, $R-S$, $\left.S-Q \right\rbrace$ is $\frac{1}{\sqrt{3}}$ times integers.

First, case (I\hspace{-.1em}I\hspace{-.1em}I) is treated. The other cases are provided in Appendix \ref{removable} as limiting values of the case (I\hspace{-.1em}I\hspace{-.1em}I) result.

The last factor of Equation (\ref{grid}) for positive $\sigma_1$+$\sigma_2$+$\sigma_3$ can be calculated as follows (see Appendix \ref{calcal}):
\begin{eqnarray}
&&\lim_{T,U \to \infty}\frac{1}{T}\int _{-\infty}^{\infty}\!\!\!  dz \Pi^3(z)h(z) \nonumber \\
&=&-i \sum_{C\left\lbrace Q,R,S\right\rbrace} 
\frac{e^{-(2N+1)\sqrt{3}\pi i(\sigma_1 +\sigma_2 +\sigma_3)Q}}{\sin\left(\sqrt{3}\pi (Q-R)\right)\sin\left(\sqrt{3}\pi (S-Q)\right)} .
\label{gg}
\end{eqnarray}
Then, multiplying $-1$ to the right-hand side of Equation (\ref{gg}) gives the correct equation for negative $\sigma_1 +\sigma_2 +\sigma_3$.

As in Section \ref{one}, to calculate $v_g(\alpha,\beta)$,
\begin{equation}
\frac{\sqrt{3}}{2(2 i)^{3}} \sigma_1 \sigma_2 \sigma_3 e^{2\pi i (\sigma_1 Q+\sigma_2 R+\sigma_3 S)}\lim_{T,U \to \infty}\frac{1}{T}\int _{-\infty}^{\infty}\!\!\!  dz \Pi^3(z) h(z)
\end{equation}
in Equation (\ref{grid}) is summed up with the use of Table \ref{sign}. 
 The result is as follows:
\begin{eqnarray}
v_g(\alpha,\beta)&=& 2\pi^2 \gamma \sum_{C\left\lbrace q,r,s\right\rbrace}\frac{\cos(3(2N+1)\pi q)}{\sin(3\pi s)\sin(3\pi r)}.
\label{result2}
\end{eqnarray}
This is the diffraction amplitude for a hexagonally-truncated triangular grid function. 
  
The previously published expression \citep{zeiders1998diffraction} is equivalent to Equation (\ref{result2}). The proof uses the equation, $\sin(u)\sin(v)=\frac{1}{2}\left(\cos(u-v)-\cos(u+v)\right)$.
\subsection{Combined Expression}
\label{three}
As shown in Equations (\ref{FP1}) and (\ref{FP2}), 
the product of Equations (\ref{result1}) and (\ref{result2}) provides the following expression for the diffraction amplitude of a hexagonally segmented telescope:
\begin{eqnarray}
v_c(\alpha,\beta)&=&4 \pi^4 \gamma^2  \sum_{C\left\lbrace Q,R,S\right\rbrace}\frac{\cos(2(1-\Delta)\pi\sqrt{3}Q)}{(\pi\sqrt{3}R)(\pi\sqrt{3}S)} \nonumber \\
&\times& \sum_{C\left\lbrace q,r,s\right\rbrace}\frac{\cos(3(2N+1)\pi q)}{\sin(3\pi s)\sin(3\pi r)},
\label{resultlast}  
\end{eqnarray}
where the variables, $q$, $r$, $s$, $Q$, $R$, and $S$ are defined in Equations (\ref{qrs}) and (\ref{QRS}); $\Delta$ denotes $\frac{2}{\sqrt{3}}$ times the half width of the gap between adjacent segments ($1-\Delta$ corresponds to the side length of a segment.); $\sqrt{3}$ is the distance between the centres of adjacent hexagons; $\gamma$ is defined in Equation (\ref{gamma}), and $\sum_{C\left\lbrace x_1,x_2,x_3\right\rbrace}$ indicates the sum over the cyclic permutations of $\left\lbrace x_1,x_2,x_3\right\rbrace$.    
This expression has removable singularities (see Appendix \ref{removable}). 

It is easy to recognize that the functions in this expression are symmetric against any symmetric operation of a regular hexagon (Section \ref{3var}.).

It can also be recognized that the function in Equation (\ref{result2}) is similar to that of Equation (\ref{result1}). This similarity is the result of the symmetry of the functions and is clearly different in nature from the expressions in previous works \citep{smith1974diffraction,nelson1985design,chanan1999strehl,zeiders1998diffraction}. The periodicity of the denominator in Equation (\ref{result2}) makes this function periodic like the Dirichlet kernel of the diffraction-grating theory \citep[e.g.][]{born2000principles}.
 In the neighborhood of the point where $q$, $r$, and $s$ are $\frac{1}{3}$ times integers, the sine function of the denominator in Equation (\ref{result2}) can be approximated by using the equation, $\lim_{x\to \pi j} \frac{\sin(x)}{x-\pi j}=(-1)^j$ ($j$ is an integer.). Therefore, Equation (\ref{result2}) has diffraction peaks located at points, $q=\frac{i}{3}$, $r=\frac{j}{3}$, and $s=\frac{k}{3}$ ($i$, $j$, and $k$ are integers,  $i+j+k=0$). These integers, $i$,$j$, and $k$, correspond to diffraction orders. Then, the profile of each peak is $\sqrt{3}\left(N+\frac{1}{2}\right)$ times sharper than that in Equation (\ref{result1}) and rotated by $\frac{\pi}{2}$ radians. This nature was not revealed directly in previously published expressions \citep{zeiders1998diffraction} where the regular-triangular symmetry was not focused.
  
The former factor of Equation (\ref{resultlast}), which is same as Equation (\ref{result1}), is the diffraction amplitude of a single hexagonal aperture with a side-length of $1-\Delta$, and this factor works just like the diffraction efficiency of a diffraction grating. The higher order peaks (i.e. the larger $|i|$,$|j|$, and $|k|$) are suppressed by this factor. When $\Delta\ll1$, the centres of the diffraction peaks become almost zero except for the central point. This is because the centres of the diffraction peaks, $Q=\frac{j-k}{\sqrt{3}}$, $R=\frac{k-i}{\sqrt{3}}$, and $S=\frac{i-j}{\sqrt{3}}$, are located closely to the points where the diffraction efficiency factor of Equation (\ref{resultlast}) becomes zero, $Q=\frac{j-k}{\sqrt{3}(1-\Delta)}$, $R=\frac{k-i}{\sqrt{3}(1-\Delta)}$, and $S=\frac{i-j}{\sqrt{3}(1-\Delta)}$. Examples of Equation (\ref{resultlast}) results are in Appendix \ref{Ex}.
\section{Discussion}
 In this paper, it must be noted that the following assumptions or idealizations are made to derive an simple analytical expression for PSFs of hexagonally segmented telescopes.  (1) Segmented mirrors are packed to make a hexagonal telescope as a whole. (2) Segment patterns are given along surfaces of primary mirror with finite radius of curvature. Thus, in fact, segment patterns projected onto  pupil planes are distorted from the patterns used in this  paper. (3) The effects of obscuration by secondary-mirror units and spiders are ignored.

In this section, how to evaluate diffraction amplitudes with additional considerations of these effects is discussed.  
　
\subsection{Factors To Be Evaluated Analytically}
\subsubsection{Segment Arrangement}
In the case of TMT or ELT, segmented mirrors are packed to make an approximately circular telescope as a whole.
These pupil functions are obtained by removing the segments near the corners of the hexagon from the arrangement in which the segmented mirrors are packed to make a hexagonal telescope. Thus, the corresponding diffraction amplitudes are obtained by subtracting those corresponding to the removed segments from the original amplitudes. When the six segments with centre points of $(A,B,C)=(\sqrt{3}K,\sqrt{3}L,\sqrt{3}M)$ ($ K+L+M = 0$) and those expressed by the sign inversion and/or cyclic permutations of $(K,L,M)$ are to be removed, subtracting 
\begin{equation}
v_u(\alpha,\beta )\sum_{C_{\left\lbrace K,L,M\right\rbrace}}2 \cos (2\pi \sqrt{3}(QK+RL+SM))
\end{equation}
from $v_c(\alpha,\beta )$ is required; This removal has six-fold rotational symmetry.
Also, central obscuration by secondary mirror units is expressed by removing the segments near the central region. 
For example, in the case of $N=3$, the Keck-type arrangement is obtained by removing central segment, $(K,L,M)=(0,0,0)$.
The TMT-type arrangement is the case of $N=13$, and the segments to be removed are expressed by 
\begin{eqnarray}
 &&(K,L,M)\nonumber \\
&=&(0,0,0),(1,-1,0),\nonumber \\
 &&(12,-12,0),(13,-13,0), \nonumber \\
 && (13-j,-13,j), (13,-13+j,-j)
\end{eqnarray}
,  and their sign inversion and/or cyclic permutations, where $j=1,2,3$.
The ELT-type arrangement is the case of $N=21$. The centre segments in the region of $|K|,|L|,|M|\leq 5$ except for
\begin{eqnarray} 
&&(K,L,M) \nonumber\\
&=&(5,-5,0),(-5,5,0),(0,5,-5),\nonumber \\
&&(0,5,-5),(-5,0,5), (5,0,-5)   
\end{eqnarray}
 are removed. The corner segments to be removed are expressed by
\begin{eqnarray} 
&&(K,L,M) \nonumber\\
&=&(19,-19,0),  (20,-20,0), (21,-21,0),\nonumber \\
&& (20-k,-20,k), (20,-20+k,-k),\nonumber \\
&& (21-l,-21,l), (21,-21+l,-l)   
\end{eqnarray}
,  and their sign inversion and/or cyclic permutations, where $k=1,2$ and $l=1,2,...,6$.
PSFs of Keck-type, TMT-type, and ELT-type pupils evaluated by these removals are shown in Appendix \ref{Ex}.
\subsection{Factors To Be Evaluated Numerically}
\subsubsection{Segment-pattern Distortion }
Segments are made to be regular hexagons along the surface of primary mirrors. Hence, projected patterns of  these onto pupil planes  are distorted from the ones considered so far in this paper. By replacing  $(x,y)$ with  $\left(R\arcsin\left(\frac{x}{R}\right)\right.$, $\left.R\arcsin\left(\frac{y}{R}\right)\right)$, this effect can be considered. Here, $R$ is the radius of curvature of the primary mirror normalized by the side length of the segments. Numerical calculations are required for investigation of the effect of the distortion on the PSFs.
\subsubsection{Spider}
Since real spiders are complex structures, accurate evaluation of effects on diffraction amplitudes have to be done by careful numerical calculations.
However, by using the variables used in this paper, a simplified spider extending  in six directions radially from the center with the width of $w$ times segment side length can be expressed simply; the values of pupil functions in the region where any one of $|A|$, $|B|$, or $|C|$ is equal to or less than $\frac{w}{2}$ is set to zero. This effect on the PSFs can be considered by numerical calculations. 
\section{Conclusion}
\label{conc}
By using three variables  (Subsection \ref{3var}), the Fourier transform of hexagonal aperture functions were related to the 3-D Fourier transform of cubic pupil functions (Subsection \ref{FF}). 
These variables were chosen so the permutations of the three variables would correspond to the permutations of the regular triangle vertices. For regular triangles, permutations of vertices are symmetry operations. Thus, the functions in resultant expression have highly obvious regular-triangular symmetry. The resultant diffraction amplitude of a regular hexagonal aperture is the Fourier transform of the 3-D equilateral rectangular function integrated along a line perpendicular to the plane on which the sum of the three variables is zero (see Figures \ref{abc3d} and \ref{qrssinc}). The diffraction amplitude of the hexagonally-truncated triangular grid function is the Fourier transform of the cubically-truncated 3-D grid function integrated in the same manner.
The diffraction amplitudes of the unit regular hexagonal aperture and hexagonally-truncated triangular grid function were derived in the same manner (Section \ref{one} and \ref{two}). Thus, each function in resultant expressions resembles the other in form. 
The remaining difference between them corresponds to the difference between sinc function and Dirichlet kernel used in basic diffraction grating theory\citep{howell2001principles}. The new expression directly shows that hexagonally segmented telescopes are diffraction gratings.  
%%%%%%%%%%%%%%%%% BODY OF PAPER %%%%%%%%%%%%%%%%%%

\section*{Acknowledgements}

The anonymous reviewer has given valuable comments on this manuscript. Allow us to express our deep gratitude to the review. The authors would like to thank Enago (www.enago.jp) for the English language review. 
%%%%%%%%%%%%%%%%%%%%%%%%%%%%%%%%%%%%%%%%%%%%%%%%%%

\bibliographystyle{mnras.bst}
\bibliography{BIBBIB} % if your bibtex file is called example.bib
%%%%%%%%%%%%%%%%%%%%%%%%%%%%%%%%%%%%%%%%%%%%%%%%%%

%%%%%%%%%%%%%%%%% APPENDICES %%%%%%%%%%%%%%%%%%%%%
\appendix

\section{permutation of two variables and symmetry operations of regular triangle}
\label{impossibility}
Two variables, $X$ and $Y$, are defined using Cartesian coordinates, $x$ and $y$.
\begin{eqnarray}
X&=& \xi (x,y) \nonumber \\
Y&=& \eta (x,y)  .
\end{eqnarray}
Here, in order to use $(X,Y)$ to express the position in 2-D plane, the mapping $f: (x,y) \to (X,Y)$ must have one-to-one correspondence. 
An operation that is not identity mapping is considered and written as $\Upsilon: (x,y) \to (x',y')$. 
Assuming that the permutation of $\left\lbrace x,y\right\rbrace$ corresponds to the operation, $\Upsilon$, Equations (\ref{ape1}) must be satisfied:
\begin{eqnarray}
X&=& \xi (x,y)= \eta (\Upsilon(x,y)) \nonumber \\
Y&=& \eta (x,y)=\xi(\Upsilon(x,y)) .
\label{ape1}
\end{eqnarray}
Then, Equations (\ref{ape2}) are also satisfied:
\begin{eqnarray}
\xi (x,y)&=& \xi (\Upsilon(\Upsilon(x,y))) \nonumber \\
\eta (x,y)&=& \eta(\Upsilon(\Upsilon(x,y))).
\label{ape2}
\end{eqnarray}
Thus,  $\Upsilon \circ \Upsilon $ must be identity mapping so as to satisfy the condition that the mapping $f: (x,y) \to (X,Y)$ must have one-to-one correspondence. For example, $R\left(\frac{2\pi}{3}\right) \circ R\left(\frac{2\pi}{3}\right)$ is not identity mapping, where $R\left(\frac{2\pi}{3}\right)$ indicates a rotational transformation by $\frac{2\pi}{3}$ rad. Hence, $\Upsilon$ cannot take $R\left(\frac{2\pi}{3}\right)$. Consequently, the symmetric operations of the regular triangle cannot be represented by the permutations of two variables.
\section{Delta function}
\label{delta}
Consider the following improper integral:
\begin{equation}
\int^{\infty}_{-\infty}d\alpha e^{-2\pi i \alpha(x-k)}=\lim_{t\to\infty,t'\to\infty}\left\lbrace\left(\int^{0}_{-t}+\int^{t'}_{0}\right)d\alpha e^{-2\pi i \alpha(x-k)}\right\rbrace.
\end{equation}
In the operation, $\lim_{t\to\infty,t'\to\infty}$, it is assumed that $t=t'$ in order to determine the improper integral value. Hence, the following equation is valid:
\begin{eqnarray}
\int^{\infty}_{-\infty}d\alpha e^{-2\pi i \alpha(x-k)}&=&\lim_{t\to\infty}\int^{t}_{-t}d\alpha e^{-2\pi i \alpha(x-k)}\nonumber \\
&=&\lim_{t\to\infty}\frac{\sin{2\pi t(x-k)}}{\pi(x-k)} \nonumber\\
&=& \delta(x-k).
\label{valid1}
\end{eqnarray}

Then, according to the theory of Fourier inverse transform, 
\begin{eqnarray} 
&\int^{\infty}_{-\infty}&\!\!\!\!\!d\alpha e^{-2\pi i \alpha x}\int^{\infty}_{-\infty}dx' f(x') e^{-2\pi i \alpha x'} \nonumber\\ &=& \frac{1}{2}\left( \lim_{\epsilon \to +0}(f(x+\epsilon)+f(x-\epsilon))\right) 
\label{vvv}
\end{eqnarray}
is valid for a piecewise smooth and square-integrable function, $f(x)$.
Equation (\ref{vvv}) becomes the following equations:
\begin{eqnarray}
&\int^{\infty}_{-\infty}&\!\!\!\!\!dx' \int^{\infty}_{-\infty}d\alpha e^{-2\pi i \alpha (x-x')}f(x') \nonumber\\ &=& \frac{1}{2}\left( \lim_{\epsilon \to +0}(f(x+\epsilon)+f(x-\epsilon) )\right)
\end{eqnarray}
\begin{equation}
\int^{\infty}_{-\infty}dx' \delta(x-x') f(x')  = \frac{1}{2}\left( \lim_{\epsilon \to +0}(f(x+\epsilon)+f(x-\epsilon) )\right).
\label{deltabasis}
\end{equation}

The function, $f(x')$, in Equation (\ref{deltabasis}) cannot be substituted by $e^{-2\pi i\alpha x'}$ because $e^{-2\pi i\alpha x'}$ is not square-integrable. 
Instead, assume that 
\begin{equation}
\int^{\infty}_{-\infty}dx' \delta(x-x') e^{-2\pi i\alpha x'}=\lim_{\rho \to 0}\int^{\infty}_{-\infty}dx' \delta(x-x') e^{-\rho |x'|}e^{-2\pi i\alpha x'}
\end{equation}
 where  $e^{-\rho |x'|}$ is a convergence factor. Thus, the following equation is valid:
\begin{equation}
\int^{\infty}_{-\infty}dx' \delta(x-x') e^{-2\pi i\alpha x'}=e^{-2\pi i\alpha x}.
\label{valid2}
\end{equation}

Equations (\ref{valid1}) and (\ref{valid2}) can be regarded as the Fourier transforms of $e^{-2\pi i \alpha k}$ and $\delta(x-x')$, respectively, in the extended meaning.
\section{Removable singularities}
\label{removable}
A function, $f(x,y,z)$  ($x$$+$$y$$+$$z$$=$$0$) may have a singularity at $(x,y,z)$$=$$(x_s,y_s,z_s)$.  When the limit value of the function, $\lim_{(x,y,z)\to(x_s,y_s,z_s)}f(x,y,z)$, is determined uniquely, the singularity is removable by redefining $\lim_{(x,y,z)\to(x_s,y_s,z_s)}f(x,y,z)$ as $f(x_s,y_s,z_s)$. 

Here, Equation (\ref{single}) is evaluated in cases (i) and (i\hspace{-.1em}i) as a limit value of the result for case (i\hspace{-.1em}i\hspace{-.1em}i), Equation (\ref{result1}). Similarly, Equation (\ref{grid}) is evaluated in cases (I) and (I\hspace{-.1em}I) as a limit value of the result for case (I\hspace{-.1em}I\hspace{-.1em}I), Equation (\ref{result2}).

\subsection{Case (i)}
The following function is defined:
\begin{equation}
hsinc(x_1,x_2,x_3)=\sum_{C\left\lbrace x_1,x_2,x_3\right\rbrace} \frac{\cos(2\pi x_1)}{\pi x_2 \pi x_3},
\label{C1}
\end{equation}
where $\sum_{C\left\lbrace x_1,x_2,x_3\right\rbrace}$ indicates the summation over the cyclic permutations of $\left\lbrace x_1,x_2,x_3\right\rbrace$.
By using Equation (\ref{C1}), Equation (\ref{result1}) can be written as
\begin{equation}
v(\alpha,\beta)=\sqrt{3}(2i)^{-2}hsinc(\sqrt{3}Q,\sqrt{3}R,\sqrt{3}S).
\end{equation}
Then, by using the l'H$\mathrm{\hat{o}}$pital's rule, the following limit value is calculated:
\begin{eqnarray}
&\lim_{\epsilon \to 0}&hsinc(\epsilon,f\epsilon,g\epsilon)\ \ \ \ \ (f+g=-1)\nonumber \\
&=&\pi^{-2}\lim_{\epsilon \to 0}\frac{\cos (2\pi \epsilon )+f\cos (2\pi f\epsilon )+g\cos (2\pi g\epsilon )}{fg\epsilon^2} \nonumber \\
&=&\pi^{-2}\lim_{\epsilon \to 0}\frac{-2\pi(\sin (2\pi \epsilon )+f^2\sin (2\pi f\epsilon )+g^2\sin (2\pi g\epsilon ))}{2fg\epsilon} \nonumber \\
&=&\pi^{-2}\lim_{\epsilon \to 0}\frac{-4\pi^2(\cos (2\pi \epsilon )+f^3\cos (2\pi f\epsilon )+g^3\cos (2\pi g\epsilon ))}{2fg} \nonumber \\
&=&\frac{-2(1+f^3+g^3)}{fg}\nonumber \\
&=&\frac{-2(3fg)}{fg}\nonumber \\
&=&-6.
\end{eqnarray}

Thus, the following equation is obtained:
\begin{eqnarray}
\lim_{Q\to0,R\to0,S\to0}v(\alpha,\beta)&=&\sqrt{3}(2i)^{-2}\times -6 \nonumber \\
&=& \frac{3\sqrt{3}}{2}.
\end{eqnarray}
This limit value does not depend on the value of $f$ and $g$. Thus, the singularity at $(Q,R,S)=(0,0,0)$ is properly removed with this limit value. 

\subsection{Case (i\hspace{-.1em}i)}
Now, by using the l'H$\mathrm{\hat{o}}$pital's rule, the following limit value is calculated:
\begin{eqnarray}
&\lim_{\epsilon \to 0}&hsinc(\epsilon,a-f\epsilon,-a-g\epsilon) \nonumber\\
&=&\pi^{-2}\left(\lim_{\epsilon \to 0}\frac{1}{\epsilon}\left\lbrace\frac{\cos{(2\pi(a-\epsilon f))}}{-a-\epsilon g}+\frac{\cos{(2\pi(-a-\epsilon  g))}}{a-\epsilon f}\right\rbrace\right.\nonumber\\
& &\left.-\frac{1}{a^2}\right)\nonumber\\
\end{eqnarray}
\begin{eqnarray}
&=&\pi^{-2}\left(\lim_{\epsilon \to 0} \left\lbrace\frac{g\cos(2\pi(a-\epsilon f))}{(a+\epsilon g)^2}-\frac{2\pi f\sin(2\pi(a-\epsilon  f))}{a+\epsilon g}\right.\right.\nonumber\\
& & \left. +\frac{f\cos(2\pi(a+\epsilon g))}{(a-\epsilon f)^2}-\frac{2\pi g\sin(2\pi(a+\epsilon g))}{a-\epsilon f}\right\rbrace\nonumber\\
& &\left.-\frac{1}{a^2}\right)\nonumber\\
\end{eqnarray}
\begin{eqnarray}
&=&\pi^{-2}\left(\frac{g \cos(2\pi a)}{a^2}-\frac{2\pi f\sin(2\pi a)}{a}\right.\nonumber\\
& & \left.+\frac{f\cos(2\pi a)}{a^2}-\frac{2\pi g\sin(2\pi a)}{a} -\frac{1}{a^2}\right)\nonumber\\
&=&\frac{\cos(2\pi a)-1}{(\pi a)^2}-\frac{2 \sin(2\pi a)}{ \pi a}.
\end{eqnarray} 
Thus, the following equation is obtained:
\begin{eqnarray}
&&\lim_{Q\to 0,R\to R,S\to -R}v(\alpha,\beta )\nonumber\\
&=&2\pi^2 \gamma\left(\frac{\cos(2\pi \sqrt{3}R)-1}{(\pi \sqrt{3}R)^2}-\frac{2 \sin(2\pi \sqrt{3}R)}{ \pi \sqrt{3}R}\right). \nonumber \\
\end{eqnarray}
This limit value does not depend on  $f$ and $g$. Thus, the singularities at $(Q,R,S)=(0,R,-R)$ are properly removed with this limit value. 

\subsection{Case (I)}
\begin{equation}
hD_{N}(x_1,x_2,x_3)=\sum_{C\left\lbrace x_1,x_2,x_3\right\rbrace} \frac{\cos\left(2\pi \left(N+\frac{1}{2}\right) x_1\right)}{\pi x_2 \pi x_3}.
\label{C9}
\end{equation}
By using Equation (\ref{C9}), Equation (\ref{result2}) can be written as 
\begin{equation}
v(\alpha,\beta)=\sqrt{3}(2i)^{-2}hD_{N}(3q,3r,3s).
\end{equation}
Then, the following limit value is calculated ($k,l,m$ are integers; $k+l+m=0$.):
\begin{eqnarray}
&\lim_{\epsilon \to 0}&hD_{N}(\epsilon+k,f\epsilon+l,g\epsilon+m)\ \ \ \ \ (f+g=-1)\nonumber \\
&=&\left(N+\frac{1}{2}\right)^{2}\lim_{\epsilon \to 0}hsinc(\epsilon,f\epsilon,g\epsilon)\ \ \ \ \ (f+g=-1)\nonumber \\
&=&-6\left(N+\frac{1}{2}\right)^{2}.
\end{eqnarray}
Thus, the following equation is obtained:
\begin{eqnarray}
\lim_{3q\to k,3r\to l,3s \to m}v(\alpha,\beta)&=&\sqrt{3}(2i)^{-2}\times -6\left(N+\frac{1}{2}\right)^{2} \nonumber \\
&=& \frac{3\sqrt{3}}{2}\left(N+\frac{1}{2}\right)^{2}.
\end{eqnarray}
This limit value does not depend on $f$ and $g$. Thus, the singularities at $(q,r,s)=\left(\frac{k}{3},\frac{l}{3},\frac{m}{3}\right)$ are properly removed with this limit value.  

\subsection{Case (I\hspace{-.1em}I)}
Now, by using the l'H$\mathrm{\hat{o}}$pital's rule, the following limit value is calculated ($k,l,m$ are integers; $k+l+m=0$; $a$ is not an integer.):
\begin{eqnarray}
&\lim_{\epsilon \to 0}&hD_{N}(\epsilon+k,a-f\epsilon+l,-a-g\epsilon+m) \nonumber\\
&=&\lim_{\epsilon \to 0}hD_{N}(\epsilon,a-f\epsilon,-a-g\epsilon) \nonumber\\
\end{eqnarray}
\begin{eqnarray}
&=&\pi^{-1}\left(\lim_{\epsilon \to 0}\frac{1}{\epsilon}\left\lbrace\frac{\cos{\left(2\pi \left(N+\frac{1}{2}\right)(a-\epsilon f)\right)}}{\sin(\pi(-a-\epsilon g))} \right.\right.\nonumber\\
&+&\left.\left.\frac{\cos{\left(2\pi\left(N+\frac{1}{2}\right)(-a-\epsilon  g)\right)}}{\sin(\pi(a-\epsilon f))}\right\rbrace
-\frac{1}{\sin ^2(\pi a)}\right)\nonumber\\
\end{eqnarray}
\begin{eqnarray}
&=&\pi^{-1}\left( \lim_{\epsilon \to 0}\left\lbrace \frac{g\cos(\pi(a+\epsilon g))\cos{\left(2\pi \left(N+\frac{1}{2}\right)(a-\epsilon f)\right)}}{\sin^2(\pi(a+\epsilon g))}\right.\right.\nonumber\\
&-&\frac{2\pi f\sin\left(2\pi\left(N+\frac{1}{2}\right)(a-\epsilon  f)\right)}{\sin(\pi(a+\epsilon g))}\nonumber\\
&+&  \frac{f\cos(\pi(a-\epsilon f)) \cos{\left(2\pi\left(N+\frac{1}{2}\right)(a+\epsilon  g)\right)}}{\sin^2(\pi(a-\epsilon f))} \nonumber\\
&-& \left. \frac{2\pi g\sin\left(2\pi\left(N+\frac{1}{2}\right)(a+\epsilon g)\right)}{\sin(\pi(a-\epsilon f))}\right\rbrace\nonumber\\
& &\left.-\frac{1}{\sin ^2(\pi a)}\right)\nonumber\\
\end{eqnarray}
\begin{eqnarray}
&=&\pi^{-1}\left(\frac{g\cos(\pi a)\cos(2\pi \left(N+\frac{1}{2}\right) a)}{\sin ^2(\pi a)}-\frac{2\pi f\sin(2\pi \left(N+\frac{1}{2}\right) a)}{\sin (\pi a)}\right.\nonumber\\
&+&\!\!\!\!\!\! \left.\frac{f\cos(\pi a)\cos \left(2\pi \left(N+\frac{1}{2}\right)a\right)-1}{\sin ^2(\pi a)}-\frac{2\pi g\sin(2\pi \left(N+\frac{1}{2}\right) a)}{\sin (\pi a)} \right)\nonumber\\
&=&\!\!\!\!\!\! \frac{\cos(\pi a)\cos \left(2\pi \left(N+\frac{1}{2}\right)a\right)-1}{\sin ^2(\pi a)}-\frac{2\pi \sin(2\pi \left(N+\frac{1}{2}\right) a)}{\sin (\pi a)}. \!\!\!\!\!\! 
\end{eqnarray}
Thus, the following equation is obtained: 
\begin{eqnarray}
&&\lim_{3q\to k,3r\to 3r+l,3s\to -3r+m}v(\alpha,\beta)\nonumber \\
&=&\!\!\!\!\!\! 2\pi^{2}\gamma \left(\frac{\cos(\pi 3r)\cos \left(2\pi \left(N+\frac{1}{2}\right)3r\right)-1}{\sin ^2(\pi 3r)}-\frac{2\pi \sin(2\pi \left(N+\frac{1}{2}\right) 3r)}{\sin (\pi 3r)}\right). \nonumber\\
\end{eqnarray}
This limit value does not depend on the values of $f$ and $g$. Thus, the singularity at $(q,r,s)=\left(\frac{k}{3},r+\frac{l}{3},-r+\frac{m}{3}\right)$ is properly removed with this limit value.  
\section{Part of Calculation}
\label{calcal}
\begin{figure}
\subfigure[$\sigma_1 +\sigma_2 +\sigma_3 >0$]{
\includegraphics[width=80mm,trim=6cm 16.5cm 8.5cm 5cm]{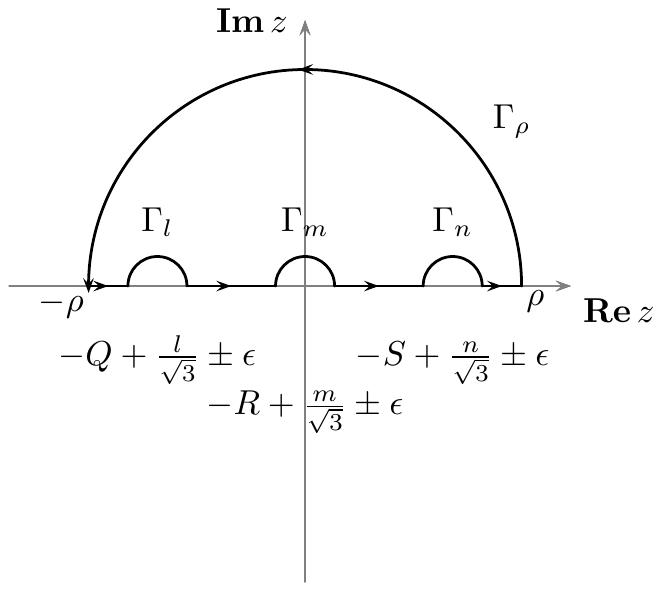}
    }
\subfigure[$\sigma_1 +\sigma_2 +\sigma_3 <0$]{
\includegraphics[width=80mm,trim=6cm 16.5cm 8.5cm 5cm]{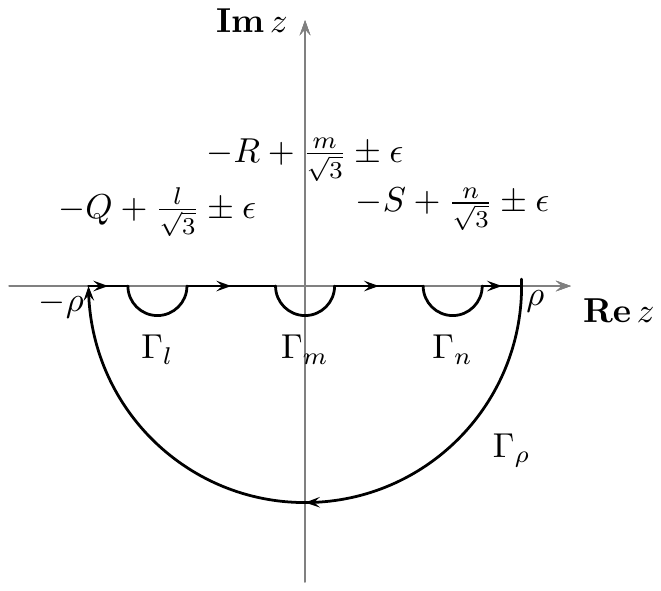}
}   
\caption{\label{integralpath2}(a) Integral path in case (I\hspace{-.1em}I\hspace{-.1em}I) for positive $\sigma_1$$+$$\sigma_2$$+$$\sigma_3$. (b) Integral path in case (I\hspace{-.1em}I\hspace{-.1em}I) for negative $\sigma_1$$+$$\sigma_2$$+$$\sigma_3$. The symbols, $l$, $m$, and $n$ are arbitrary integers. The integrand must be regular inside the integral paths.}
\end{figure} 
An integral path shown in Figure \ref{integralpath2} is considered for $\Pi^3(z)h(z)$ in Equation (\ref{grid}). The outer semicircle is denoted by $\Gamma_\rho$ and the semicircles for avoiding singularities are denoted by $\Gamma_l$, $\Gamma_m$, and $\Gamma_n$. The radius of $\Gamma_\rho$ approaches infinity ($\rho\to\infty$); and the radii of $\Gamma_l$, $\Gamma_m$, and $\Gamma_n$ approach zero ($\epsilon\to0$).
By applying the Cauchy integral theorem along the integral path, the last factor of Equation (\ref{grid}) can be calculated as follows: 
\begin{equation}
\int _{-\infty}^{\infty}\!\!\!  dz \Pi^3(z)h(z) = -\sum_{C\left\lbrace l,m,n\right\rbrace}\sum_{l=-\infty}^{\infty} \int_{\Gamma_{l}} dz \Pi^3(z)h(z).
\label{gamm2}
\end{equation}
Hence, when $\sigma_1+\sigma_2+\sigma_3$ is positive, 
\begin{equation}
\lim_{T,U \to \infty}\frac{1}{T}\int _{-\infty}^{\infty}\!\!\!  dz \Pi^3(z)h(z) \nonumber
\end{equation}
becomes
\begin{eqnarray}
&=&-\lim_{T,U \to \infty}\frac{1}{T}\lim_{\epsilon \to 0}\int_{\pi}^0 i\epsilon e^{i\theta} d\theta \nonumber  \\
&&\sum_{C\left\lbrace Q,R,S\right\rbrace}\left\lbrace \sum_{l=-\infty}^{\infty}\Pi^3\left(\epsilon e^{i \theta}-Q+\frac{l}{\sqrt{3}}\right)h\left(\epsilon e^{i \theta}-Q+\frac{l}{\sqrt{3}}\right)\right\rbrace \nonumber \\
&=&-\sum_{C\left\lbrace Q,R,S\right\rbrace}\left\lbrace \lim_{T \to \infty}\frac{1}{T}\sum_{l=-\infty}^{\infty}\lim_{U \to \infty}\Pi^3\left(-Q+\frac{l}{\sqrt{3}}\right) \right.\nonumber\\
&&\left. \lim_{\epsilon \to 0} \int_{\pi}^0 i\epsilon e^{i\theta} d\theta h\left(\epsilon e^{i \theta}-Q\right) \right\rbrace \nonumber%\\
\end{eqnarray}
\begin{eqnarray}
&=& -\sum_{C\left\lbrace Q,R,S\right\rbrace}\left\lbrace \sum_{l=-\infty}^{\infty}\lim_{T \to \infty}\frac{1}{T}\mathrm{Rect}\left(\frac{l}{\sqrt{3}T}\right)\right.\nonumber\\ &&\left.\lim_{\epsilon \to 0}\int_{\pi}^0 i\epsilon e^{i\theta} d\theta h\left(\epsilon e^{i \theta}-Q\right) \right\rbrace \nonumber \\
&=& -\sqrt{3}\lim_{\epsilon \to 0}\int_{\pi}^0 i\epsilon e^{i\theta} d\theta \sum_{C\left\lbrace Q,R,S\right\rbrace} h\left(\epsilon e^{i \theta}-Q\right).
\label{gridgrid}
\end{eqnarray}
 
Thus,
\begin{eqnarray}
&&\lim_{T,U \to \infty}\frac{1}{T}\int _{-\infty}^{\infty}\!\!\!  dz \Pi^3(z)h(z) \nonumber \\
&=&-i \sum_{C\left\lbrace Q,R,S\right\rbrace} 
\frac{e^{-(2N+1)\sqrt{3}\pi i(\sigma_1 +\sigma_2 +\sigma_3)Q}}{\sin\left(\sqrt{3}\pi (Q-R)\right)\sin\left(\sqrt{3}\pi (S-Q)\right)}, \nonumber \\
\end{eqnarray}
where  $\lim_{|z| \to 0}\frac{\sin(z)}{z}$ was used.
\section{Examples}
\label{Ex}
\begin{figure}
\begin{center}
\includegraphics[width=80mm]{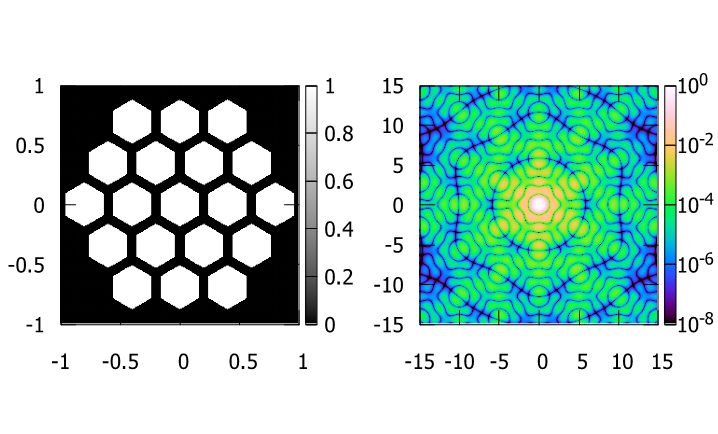}
\caption{N=2 \label{N2}}
\end{center}

\end{figure}

\begin{figure}
\begin{center}
\includegraphics[width=80mm]{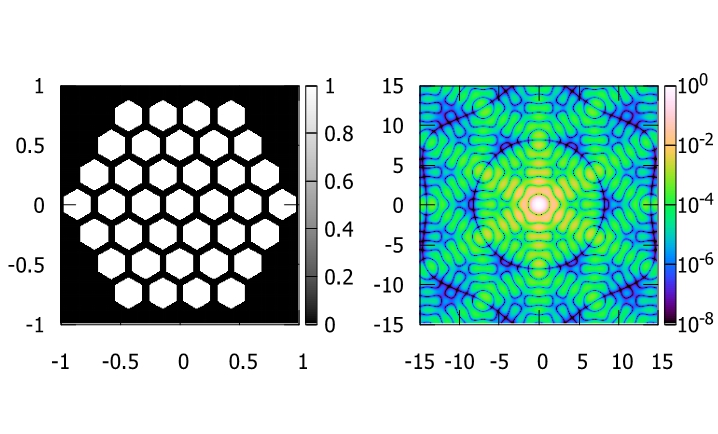}
\caption{N=3\label{N3}}
\end{center}

\end{figure}

\begin{figure}
\begin{center}
\includegraphics[width=80mm]{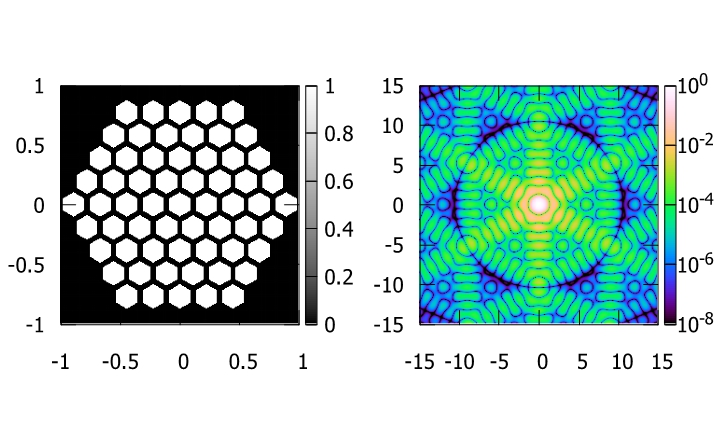}
\caption{N=4\label{N4}}
\end{center}

\end{figure}

\begin{figure}
\begin{center} 
\includegraphics[width=80mm]{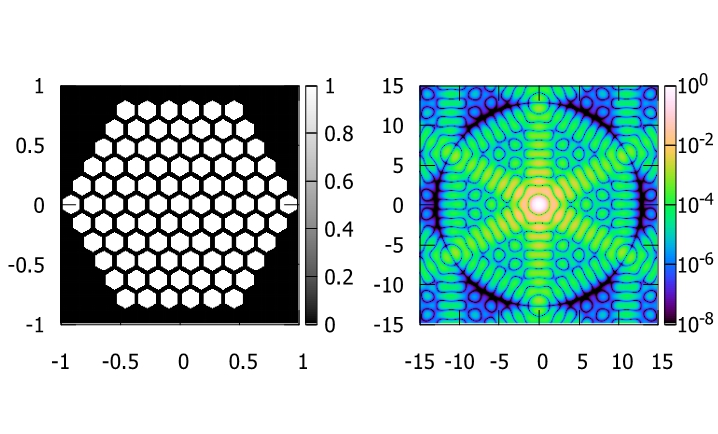}
\caption{N=5\label{N5}}
\end{center}

\end{figure}

\begin{figure}
\begin{center}
\includegraphics[width=80mm]{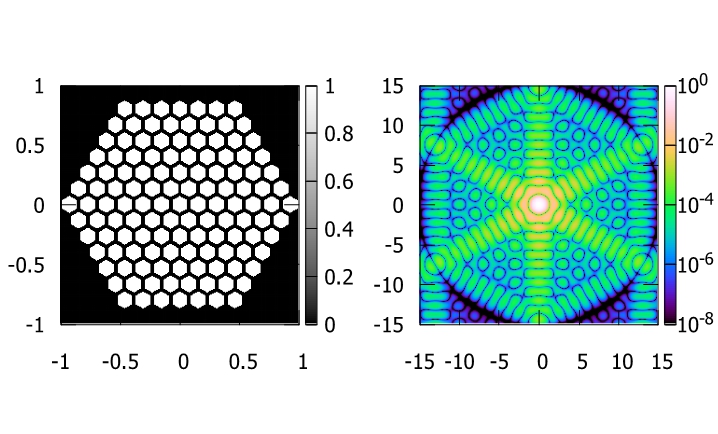}
\caption{N=6\label{N6}}
\end{center}

\end{figure}

\begin{figure}
\begin{center}
\includegraphics[width=80mm]{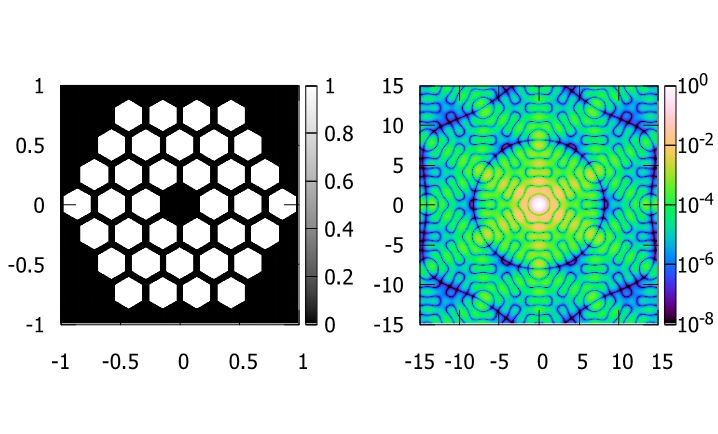}
\caption{Keck-type \label{Keck}}
\end{center}

\end{figure}

\begin{figure}
\begin{center} 
\includegraphics[width=80mm]{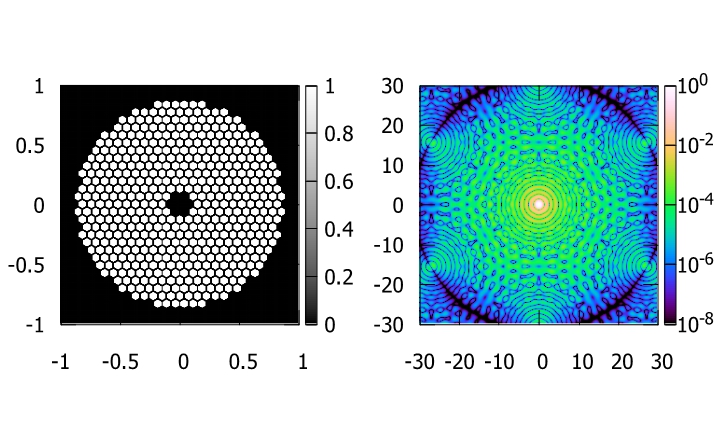}
\caption{TMT-type \label{TMT}}
\end{center}

\end{figure}

\begin{figure}
\begin{center}
\includegraphics[width=80mm]{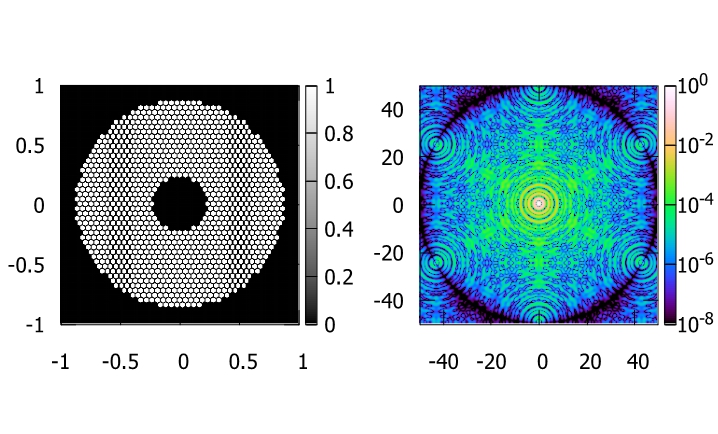}
\caption{ELT-type \label{ELT}}
\end{center}

\end{figure}

In the below PSF examples (Figure \ref{N2}--\ref{ELT}), $\Delta$ is fixed to $1.0 \times 10^{-5}$. In the figures, the gap between adjacent segments are exaggerated to become 40 times wider than that used for the present calculation. The coordinates on the focal plane are normalized by $\frac{\lambda}{D}$, where $\lambda$ is the wavelength of light, and $D$ is the full width of the pupil along the $x$-direction.

%\begin{comment}
%\end{comment}

% Don't change these lines
\bsp	% typesetting comment
\label{lastpage}
\end{document}